\documentclass[aps,prd,reprint,groupedaddress,showpacs,eqsecnum,floatfix]{revtex4-1}

\usepackage{amsmath}
\usepackage{amssymb}
\usepackage{graphicx}
\usepackage{bm}
\usepackage{braket}
\usepackage{enumitem}
\usepackage{tensor}
\usepackage{color}
\usepackage{combelow}
\usepackage{slashed}
\usepackage{color}
\usepackage{epstopdf}

\def\feq{\ensuremath{f^{(\mathrm{eq})}}}

\def\hatt{{\hat{0}}}
\def\hati{{\hat{\imath}}}
\def\hatj{{\hat{\jmath}}}
\def\hatr{{\hat{r}}}
\def\htheta{{\hat{\theta}}}
\def\hvarphi{{\hat{\varphi}}}
\def\halpha{{\hat{\alpha}}}
\def\hbeta{{\hat{\beta}}}
\def\hgamma{{\hat{\gamma}}}
\def\hrho{{\hat{\rho}}}

\def\ta{{\tilde{a}}}
\def\tb{{\tilde{b}}}
\def\tildet{{\tilde{0}}}
\def\tildei{{\tilde{\imath}}}
\def\tildej{{\tilde{\jmath}}}

\def\wK{\widetilde{K}}

\def\omegabar{\overline{\omega}}
\def\en{E}

\begin{document}

\title{Maxwell-J\"uttner distribution for rigidly-rotating flows in
spherically symmetric spacetimes using the tetrad formalism}

\author{Victor E. \surname{Ambru\cb{s}}}
\email{victor.ambrus@e-uvt.ro}
\affiliation{Department of Physics, West University of Timi\cb{s}oara,
Bd.~Vasile P\^arvan 4, Timi\cb{s}oara, 300223, Romania}
\author{Ion I. \surname{Cot\u{a}escu}}
\email{icotaescu@yahoo.com}
\affiliation{Department of Physics, West University of Timi\cb{s}oara,
Bd.~Vasile P\^arvan 4, Timi\cb{s}oara, 300223, Romania}

\date{\today}

\begin{abstract}
We consider rigidly rotating states in thermal equilibrium 
on static spherically symmetric spacetimes. 
Using the Maxwell-J\"uttner equilibrium distribution function, constructed as 
a solution of the relativistic Boltzmann equation,
the equilibrium particle flow four-vector, stress-energy tensor and the 
transport coefficients in the Marle model are computed. Their properties are 
discussed in view of the topology of the speed-of-light surface induced by the 
rotation for two classes of spacetimes: maximally symmetric (Minkowski, de Sitter 
and anti-de Sitter) and Reissner-Nordstr\"om black-hole spacetimes.
To facilitate our analysis, we employ a non-holonomic comoving tetrad 
field, obtained unambiguously by applying a Lorentz boost on a fixed 
background tetrad.
\end{abstract}


\maketitle

\section{Introduction}


Due to their simplicity, rigidly rotating systems in thermal equilibrium 
represent attractive toy-models which can be used to gain insight on 
the physical features of more complex systems or geometries which 
exhibit rotation (e.g.~rotating Kerr black holes). Such systems can be interesting 
also from a quantum field theory point of view, where the definition of vacuum 
states or states at finite temperature is still an open field (for some recent results,
see Ref.~\cite{casals13} and references therein).

On Minkowski spacetime, such systems were studied using both kinetic theory 
and quantum field theory
\cite{vilenkin78,letaw80,vilenkin80,iyer82,cercignani02,duffy03,becattini08,becattini10,
becattini11a,becattini11b,becattini13,becattini14,ambrus14a,becattini15,alba15,ambrus15} 
and the quantum corrections can be obtained analytically
\cite{becattini15,ambrus14a,ambrus15}.
In this paper, we use the relativistic Boltzmann equation to study the 
equilibrium states and the transport coefficients of fluids undergoing 
rigid rotation on static spherically-symmetric background spacetimes, 
as well as to discuss the topology of the speed of light surface (SOL) 
which forms due to the rotation. 

In order to obtain expressions for the transport coefficients, 
the Marle model is employed for the Boltzmann collision integral \cite{marle69}.
To facilitate our analysis, we employ non-holonomic tetrad fields 
\cite{lindquist66,riffert86,cardall13} with respect to which the 
mass shell condition for the momentum four-vector becomes independent 
of the background metric, while the calculation of the transport coefficients
becomes identical with that on the Minkowski spacetime \cite{cercignani02}. 
The tetrad of the comoving frame is obtained by applying a pure Lorentz boost 
(i.e.~without rotation) on the tetrad of the background metric \cite{misner73}.
The only degrees of freedom available in this procedure 
correspond to choosing the gauge for the fixed tetrad. 
Our formulation is sufficiently general to encompass previously studied examples, 
such as the Minkowski \cite{ambrus15} and Schwarzschild \cite{kremer13,kremer14} 
spacetimes. We specialize our results to the cases of maximally symmetric 
spacetimes (Minkowski, de Sitter and anti-de Sitter), as well as for the 
Reissner-Nordstr\"om spacetime.

In Sec.~\ref{sec:boltz}, we discuss the tetrad formalism, which we apply to obtain the 
transport coefficients in the Marle model. The construction of the comoving frame for 
rigidly rotating flows on spherically symmetric spaces is presented in Sec.~\ref{sec:sph}, 
while Sec.~\ref{sec:app} is dedicated to the discussion of rigidly rotating thermal states
on maximally symmetric and Reissner-Nordstr\"om spacetimes. Section~\ref{sec:conc} 
concludes this paper.

\section{The relativistic Boltzmann equation}\label{sec:boltz}

We start this section by presenting in Subsec.~\ref{sec:boltz:comoving} a technique for 
defining the comoving frame with no unspecified degrees of freedom, which relies on a 
fixed tetrad field corresponding to the (arbitrary) background spacetime.

Subsection~\ref{sec:boltz:cons} reviews the Boltzmann equation written with respect 
to tetrad fields in conservative form, as described in Ref.~\cite{cardall13}. Details
regarding the transition from the generally covariant Boltzmann equation to the Boltzmann 
equation with respect to tetrad fields, as well as from this latter form to the 
conservative form of the Boltzmann equation, are presented in 
appendices~\ref{app:boltz_tetrad} and \ref{app:boltz_cons}, respectively.

Subsection~\ref{sec:boltz:eq} introduces the Maxwell-J\"uttner distribution for 
local thermodynamic equilibrium, as well as the conditions that the macroscopic 
particle number density, temperature and four-velocity must satisfy in order 
for the fluid to be in global thermodynamic equilibrium. 

Subsection~\ref{sec:boltz:tcoeff} ends this section with a computation of the transport 
coefficients arising when the Marle model is used for the collision integral, which are 
calculated starting from the Boltzmann equation in conservative form in a manner analogous 
to that employed on flat space \cite{cercignani02}. The expressions for the resulting 
coefficients, defined by using a covariant generalisation \cite{rezzolla13} of their flat 
spacetime definitions, are identical to those obtained in flat spacetime. 

\subsection{Comoving frame}\label{sec:boltz:comoving}

Considering a fixed spacetime having the line element:
\begin{equation}
 ds^2 = g_{\mu\nu} dx^\mu dx^\nu,
\end{equation}
an orthonormal frame $\{e_\ta\}$ can be chosen such that the metric is locally flat:
\begin{equation}
 g_{\mu\nu} e^{\mu}_\ta e^\nu_\tb = \eta_{\ta\tb},\label{eq:tetrad_tilde_gen}
\end{equation}
where $\eta_{\ta \tb} = {\rm{diag}}(-1, 1, 1, 1)$ is the metric of the Minkowskian
model of this spacetime.
The tetrad frame vectors $e^\mu_{\ta}$ uniquely determine a set of co-vectors (one-forms)
$\{\omega^\ta\}$ through \cite{misner73}:
\begin{equation}
 \braket{e_\tb, \omega^\ta} \equiv \omega^\ta_\mu e^\mu_\tb = 
 \delta^\ta{}_\tb.
\end{equation}
The choice of tetrad has $6$ degrees of freedom, due to the invariance of 
Eq.~\eqref{eq:tetrad_tilde_gen} under Lorentz transformation. However,
we consider that the tetrad $\{e_\ta\}$ is fixed in some predefined gauge, 
such that it can serve as a reference tetrad for the future development in this chapter.

A comoving frame is defined as an orthonormal frame $\{e_{\halpha}\}$ with respect to which 
the fluid four-velocity is 
\begin{equation}
 u^\halpha \equiv u^\mu \omega_\mu^\halpha = (1,0,0,0),\label{eq:umu_comoving}
\end{equation}
where $\{\omega^\halpha\}$ are the co-vectors corresponding to the tetrad vectors $\{e_{\halpha}\}$.
Eq.~\eqref{eq:umu_comoving} implies:
\begin{equation}
 e_\hatt^\mu = u^\mu.\label{eq:tetrad_hat_t}
\end{equation}
Eq.~\eqref{eq:tetrad_hat_t} reduces the number of degrees of freedom in 
Eq.~\eqref{eq:tetrad_tilde_gen} to $3$. We eliminate these degrees of freedom 
by requiring that the comoving frame $\{e_\halpha\}$
is obtained from the local frame $\{e_\ta\}$ by applying a pure Lorentz boost
$L^\ta{}_\halpha$, such that:
\begin{equation}
 e_\halpha = e_\ta L^\ta{}_\halpha.
\end{equation}
The components $L^\ta{}_\hatt$ can be obtained by contracting Eq.~\eqref{eq:tetrad_hat_t} 
with $\omega^\ta_\mu$:
\begin{equation}
 L^\ta{}_\hatt = u^\ta \equiv u^\mu \omega^\ta_\mu.
\end{equation}
The above equation fixes all three degrees of freedom of the genuine Lorentz boost $L^\ta{}_\halpha$,
which can be written as follows:
\begin{equation}
 L^\ta{}_\halpha = 
 \begin{pmatrix}
  u^\tildet & u_\tildej\\
  u^\tildei & {\displaystyle \delta^\tildei{}_\tildej + \frac{u^\tildei u_\tildej}{u^\tildet + 1}}
 \end{pmatrix}.
 \label{eq:boost}
\end{equation}
It can be checked that $L^\ta{}_\hbeta$ is indeed a pseudo-orthogonal matrix:
\begin{equation}
 \eta_{\ta\tb} L^\ta{}_\halpha L^\tb{}_\hbeta = \eta_{\halpha\hbeta}, \qquad 
 \eta^{\halpha\hbeta} L^\ta{}_\halpha L^\tb{}_\hbeta = \eta^{\ta\tb},
\end{equation}
satisfying $L^T = L$ \cite{tung85}.

\subsection{Conservative relativistic Boltzmann equation}\label{sec:boltz:cons}

The relativistic Boltzmann equation with respect to arbitrary coordinate systems on
arbitrary geometries can be written as:
\begin{equation}
 p^\mu \frac{\partial f}{\partial x^\mu} - 
 \Gamma^i{}_{\mu\nu} p^\mu p^\nu \frac{\partial f}{\partial p^i} = C[f],
 \label{eq:boltz}
\end{equation}
where $f \equiv f(x^\mu, p^i)$ is the Boltzmann distribution function, $x^\mu$ represent spacetime 
coordinates and $p^\mu = (p^0, p^i)$ are the components of the particle four-momentum vector.
The time component $p^0$ of the momentum $4$-vector is fixed by the mass-shell condition:
\begin{equation}
 g_{\mu\nu} p^\mu p^\nu = -m^2,\label{eq:mshell}
\end{equation}
where $g_{\mu\nu}$ are the components of the spacetime metric.
The connection coefficients $\Gamma^i{}_{\mu\nu}$ appearing in Eq.~\eqref{eq:boltz}
have the following expression with respect to a coordinate frame:
\begin{equation}
 \Gamma^\lambda{}_{\mu\nu} = \frac{1}{2} g^{\lambda\sigma}\left(
 g_{\sigma\mu,\nu} + g_{\sigma\nu,\mu} - g_{\mu\nu,\sigma}\right),
 \label{eq:christoffel}
\end{equation}
where a comma denotes differentiation with respect to the coordinates,
e.g.~$g_{\sigma\mu,\nu} \equiv \partial_{\nu} g_{\sigma\mu} \equiv
\frac{\partial g_{\sigma\mu}}{{\partial x^\nu}}$.

The Boltzmann equation \eqref{eq:boltz} can be expressed with respect to the tetrad components of the momentum 
vector as follows:
\begin{equation}
 p^\halpha e^{\mu}_\halpha \frac{\partial f}{\partial x^\mu} - 
 \Gamma^{\hati}{}_{\halpha\hbeta} p^\halpha p^\hbeta \frac{\partial f}{\partial p^{\hati}} = \mathcal{C}[f].
 \label{eq:boltz_tetrad}
\end{equation}
For more details on the relation between Eqs.~\eqref{eq:boltz} and \eqref{eq:boltz_tetrad}, we refer the 
reader to Appendix~\ref{app:boltz_tetrad}.
The connection coefficients $\Gamma^{\hgamma}{}_{\halpha\hbeta}$ appearing in Eq.~\eqref{eq:boltz_tetrad} 
can be obtained using:
\begin{equation}
 \Gamma^{\hgamma}{}_{\halpha\hbeta} = \eta^{\hgamma\hat{\rho}} \left(
 c_{\hrho\halpha\hbeta} + c_{\hrho\hbeta\halpha} - c_{\halpha\hbeta\hrho}\right),
 \label{eq:conn_coeff}
\end{equation}
where the Cartan coefficients can be calculated from the commutators of the tetrad vectors \cite{misner73}:
\begin{equation}
 c_{\halpha\hbeta}{}^{\hgamma} = \braket{[e_\halpha, e_\hbeta], \omega^\hgamma}.
\end{equation}

In order to derive transport equations for macroscopic quantities, we follow Ref.~\cite{cardall13} and 
express Eq.~\eqref{eq:boltz_tetrad} in conservative form:
\begin{equation}
 \frac{1}{\sqrt{-g}} \partial_\mu \left(\sqrt{-g} p^\halpha e_\halpha^\mu f\right) - 
 p^\hatt \frac{\partial}{\partial p^\hati} \left(
 \Gamma^\hati{}_{\halpha\hbeta} \frac{p^\halpha p^\hbeta}{p^\hatt} f\right) = C[f].
 \label{eq:boltz_cons}
\end{equation}
For completeness, we provide the details of the derivation of the transition from 
Eq.~\eqref{eq:boltz_tetrad} to Eq.~\eqref{eq:boltz_cons} in Appendix~\ref{app:boltz_cons}.
The form~\eqref{eq:boltz_cons} of the Boltzmann equation is particularly convenient from a numerical 
point of view, being directly amenable to finite-element or finite-volume numerical methods.
Furthermore, Eq.~\eqref{eq:boltz_cons} can be used to easily derive transport equations for the 
moments $M^{\halpha_1 \dots \halpha_{n+1}}$ of $f$:
\begin{equation}
 \nabla_{\halpha_{n+1}} T^{\halpha_1 \halpha_2 \dots \halpha_n \halpha_{n+1}} = 
 \int \frac{d^3p}{p^{\hatt}} C[f] p^{\halpha_1} \dots p^{\halpha_n},\label{eq:momn}
\end{equation}
where
\begin{equation}
 T^{\halpha_1 \halpha_2 \dots \halpha_n \halpha_{n+1}} \equiv 
 \int \frac{d^3p}{p^{\hatt}} f\, p^{\halpha_1} \dots p^{\halpha_n} p^{\halpha_{n+1}}.
\end{equation}
In particular, the conservation equation for the particle four-flow $N^\alpha \equiv T^\halpha$ 
and stress-energy tensor $T^{\halpha \hbeta}$ can be obtained from Eq.~\eqref{eq:momn}
for $n = 0$ and $n = 1$:
\begin{equation}
 \nabla_\halpha N^\halpha = 0, \qquad \nabla_\hbeta T^{\halpha\hbeta} = 0.
\end{equation}
The right hand sides of the above equations vanish since $1$ and $p^\halpha$ are
collision invariants \cite{cercignani02}, i.e.:
\begin{equation}
 \int \frac{d^3p}{p^{\hatt}} C[f] = \int \frac{d^3p}{p^{\hatt}} C[f] p^\halpha = 0.
 \label{eq:coll_inv}
\end{equation}

\subsection{Thermodynamic equilibrium}\label{sec:boltz:eq}

At local equilibrium, the collision integral $C[f]$ vanishes and
$f$ is given by \cite{cercignani02,romatschke12}:
\begin{equation}
 \feq = \frac{Z}{(2\pi)^3} 
 \left[\exp\left( -\beta\mu -\beta p^\halpha u_\halpha\right) - \varepsilon\right]^{-1},
 \label{eq:feq}
\end{equation}
where $Z$ represents the number of degrees of freedom, $\beta = 1 / T$ is the inverse local
temperature, $u_\halpha$ are the covariant components of the 
macroscopic velocity $4$-vector, and $\mu$ is the chemical potential. 
The constant $\varepsilon$ takes the values $-1$, $0$ and $1$  for the Fermi-Dirac (F-D), 
Maxwell-J\"uttner (M-J) and Bose-Einstein (B-E) distributions,
respectively. Since the equilibrium distributions corresponding to the Fermi-Dirac or 
Bose-Einstein statistics can be inferred from the M-J distribution \cite{florkowski15,ambrus15},
the focus in this paper will be on the latter distribution, which we give explicitly 
below:
\begin{equation}
 \feq = \frac{Z}{(2\pi)^3} \exp\left( \beta \mu + \beta p^\halpha u_\halpha\right). \label{eq:feq_MJ}
\end{equation}
The chemical potential $\mu$ can be eliminated in favor of the particle number density $n$, as follows 
\cite{cercignani02}:
\begin{equation}
 \feq = \frac{n \beta}{4\pi m^2 K_2(m\beta)} \exp\left(\beta p^\halpha u_\halpha\right).\label{eq:feq_MJ_n}
\end{equation}
Direct integration of Eq.~\eqref{eq:feq_MJ_n} can be employed to obtain the equilibrium 
expressions of the particle flow four-vector $N^\halpha$ and of the stress-energy tensor $T^{\halpha\hbeta}$:
\begin{subequations}
\begin{align}
 N^\halpha =& nu^\halpha, \label{eq:Nalpha_eq}\\
 T^{\halpha\hbeta} =& \en u^\halpha u^\hbeta + 
 P \Delta^{\halpha\hbeta}, \label{eq:SET_eq}
\end{align}
\end{subequations}
where $\Delta^{\halpha\hbeta}$ is the projector corresponding to the hypersurface orthogonal to $u^\halpha$:
\begin{equation}
 \Delta^{\halpha\hbeta} \equiv \eta^{\halpha\hbeta} + u^\halpha u^\hbeta.\label{eq:Delta_def}
\end{equation}
In Eq.~\eqref{eq:SET_eq}, the equilibrium energy density $\en$ and 
pressure $P$ have the following expression:
\begin{subequations}\label{eq:beta_def}
\begin{align}
 \en =& nm G(\zeta) - P, \label{eq:eps_def}\\
 P =& \frac{n}{\beta},\label{eq:P_def}
\end{align}
\end{subequations}
where the relativistic coldness $\zeta$ is defined as \cite{cercignani02,rezzolla13}:
\begin{equation}
 \zeta = m\beta \label{eq:zeta_def}
\end{equation}
and the function $G(\zeta)$ is defined in terms of modified Bessel functions 
of the third kind $K_n$ \cite{cercignani02}:
\begin{equation}
 G(\zeta) = \frac{K_3(\zeta)}{K_2(\zeta)}.\label{eq:G_def}
\end{equation}
Thus, the inverse temperature $\beta$ uniquely determines the energy density $\en$ and hydrostatic pressure 
$P$ through Eqs.~\eqref{eq:beta_def}.

It is worth noting that the trace of $T^{\halpha\hbeta}$ \eqref{eq:SET_eq} has the following form:
\begin{align}
 T^\halpha{}_{\halpha} =& -m^2 \int\frac{d^3p}{p^{\hatt}} \feq\nonumber\\
 =& -\varepsilon + 3P 
 = -nm\frac{K_1(\zeta)}{K_2(\zeta)}.\label{eq:SET_eq_trace}
\end{align}

We end this subsection by noting that when the fluid is in global thermodynamic equilibrium, 
$f = \feq$ everywhere in the spacetime. Substituting Eq.~\eqref{eq:feq_MJ} into the Boltzmann 
equation in conservative form \eqref{eq:boltz_cons} shows that $\beta\mu$ must be constant, 
while the vector field $k^\halpha = \beta u^\halpha$ must satisfy the Killing equation \cite{cercignani02}:
\begin{equation}
 \nabla_\halpha(\beta \mu) = 0, \qquad 
 k_{\halpha;\hbeta} + k_{\hbeta;\halpha} = 0.\label{eq:killing}
\end{equation}
In the above, the semicolon denotes the covariant differentiation. In Section~\ref{sec:sph}, Eqs.~\eqref{eq:killing}
will be solved for the case of rigidly rotating thermal distributions on general static spherically symmetric spacetimes.

\subsection{Transport coefficients}\label{sec:boltz:tcoeff}

In an out-of-equilibrium flow, the distribution function $f$ is generally different from $\feq$.
In the Eckart decomposition, the particle flow $4$-vector $N^\halpha \equiv T^\halpha$ and 
the stress-energy tensor can be written as:
\begin{subequations}
\begin{align}
 N^\halpha =& nu^\halpha, \label{eq:Nalpha}\\
 T^{\halpha\hbeta} =& \en u^\halpha u^\hbeta + 
 (P + \omegabar) \Delta^{\halpha\hbeta} + 2q^{(\halpha} u^{\hbeta)} + \pi^{\halpha\hbeta}, 
 \label{eq:SET}
\end{align}
\end{subequations}
where the energy density $\en$ and hydrostatic pressure $P$ define the non-equilibrium inverse 
temperature $\beta$ through Eqs.~\eqref{eq:beta_def}. The energy density $\en$, 
dynamic pressure $\omegabar$, heat flux $q^\halpha$
and pressure deviator $\pi^{\halpha\hbeta}$ can be computed from $T^{\halpha\hbeta}$ using the following
expressions:
\begin{subequations}\label{eq:SET_noneq}
\begin{align}
 \en =& u_\halpha u_\hbeta T^{\halpha\hbeta},\label{eq:eps_from_SET}\\
 P+ \omegabar =& \frac{1}{3} \Delta_{\halpha\hbeta} T^{\halpha\hbeta}, \label{eq:omegabar_def}\\
 q^\halpha =& -\Delta^\halpha{}_{\hbeta} u_{\hgamma} T^{\hbeta\hgamma}, \label{eq:q_def}\\
 \pi^{\halpha\hbeta} =& T^{<\hgamma\hrho>},\label{eq:pi_def}
\end{align}
\end{subequations}
where the notation $A^{<\hgamma\hrho>}$ refers to:
\begin{equation}
 A^{<\hgamma\hrho>} \equiv \left[\frac{1}{2}\left(\Delta^\halpha{}_{\hgamma} \Delta^\hbeta{}_{\hrho} 
 + \Delta^\halpha{}_{\hrho} \Delta^\hbeta{}_{\hgamma}\right) - 
 \frac{1}{3} \Delta^{\halpha\hbeta} \Delta_{\hgamma\hrho}\right] A^{\hgamma\hrho}.
 \label{eq:angular_def}
\end{equation}

In the hydrodynamic limit, the following relations hold for the 
dynamic pressure, pressure deviator and heat flux\cite{cercignani02,rezzolla13}:
\begin{subequations}\label{eq:transp_coef_def}
\begin{align}
 \omegabar =& -\eta\nabla_{\hgamma} u^\hgamma, \label{eq:eta_def}\\
 q^\halpha =& -\lambda \Delta^{\halpha\hbeta}\left(\nabla_\hbeta T - 
 \frac{T}{\en + P} \nabla_\hbeta P\right),\label{eq:lambda_def}\\
 \pi_{\halpha\hbeta} =& -2\mu \nabla_{<\halpha} u_{\hbeta>},\label{eq:mu_def}
\end{align}
\end{subequations}
where $T = \beta^{-1}$ and the bulk viscosity $\eta$, shear viscosity $\mu$ and thermal conductivity $\lambda$ 
are the transport coefficients which make the subject of the present subsection.

The values of the transport coefficients 
depend on the form of the collision operator $C[f]$ in the Boltzmann equation 
\eqref{eq:boltz_cons}. In general, $C[f]$ is a nonlinear integral operator which drives $f$ 
towards local thermodynamical equilibrium \cite{cercignani02,florkowski10}. 
The computation of the transport coefficients requires the 
analysis of the hydrodynamic regime of the Boltzmann equation, for the recovery of which 
there are various procedures, including: the Chapman-Enskog procedure 
\cite{marle69,anderson74b}, the Grad moments method \cite{cercignani02}
and the renormalisation group method \cite{hatta02,kunihiro06,tsumura07}. 
To illustrate the methodology for the computation of the transport coefficients, we employ 
in this section the single relaxation time models proposed by Marle \cite{marle69} and 
Anderson-Witting \cite{anderson74a}:
\begin{subequations}\label{eq:coll}
\begin{align}
 C[f]_{\rm M} =& -\frac{m}{\tau} (f - \feq), \label{eq:marle}\\
 C[f]_{\rm A-W} =& \frac{u_\halpha p^\halpha}{\tau} (f - \feq)\label{eq:AW},
\end{align}
\end{subequations}
where $\tau$ is the relaxation time. For the remainder of this section, we only consider the 
Marle collision term, with which the Boltzmann equation \eqref{eq:boltz_cons} in conservative 
form reads:
\begin{multline}
 \frac{1}{\sqrt{-g}} \partial_\mu \left(\sqrt{-g} p^\halpha e_\halpha^\mu f\right) - p^{\hatt} \frac{\partial}{\partial p^\hati} 
 \left(\Gamma^\hati{}_{\halpha\hbeta} \frac{p^\halpha p^\hbeta}{p^{\hatt}} f\right)\\
 =-\frac{m}{\tau} (f - \feq).\label{eq:boltz_cons_marle}
\end{multline}

In order for the Marle model \eqref{eq:marle} to be consistent, 
the collision invariants $1$ and $p^\halpha$ must be preserved. Replacing Eq.~\eqref{eq:marle}
in Eq.~\eqref{eq:coll_inv} gives:
\begin{subequations}
\begin{align}
 \nabla_\halpha N^\halpha =& -\frac{m}{\tau} \int \frac{d^3p}{p^{\hatt}} (f - \feq) =
 \frac{1}{m\tau} (T^{\halpha}{}_{\halpha} - T^{\halpha}_E{}_{\halpha})\label{eq:marle1}\\
 \nabla_\hbeta T^{\halpha\hbeta} =& -\frac{m}{\tau} (N^\halpha - N^\halpha_E). \label{eq:marle2}
\end{align}
\end{subequations}
The above equations can be used to determine the parameters $n_E$, $u^\halpha_E$ and $T_E$ of the 
Maxwell-J\"uttner distribution $\feq$, as well as of the corresponding ``equilibrium'' stress-energy 
tensor $T^{\halpha\hbeta}_E$. Since $N^\halpha = nu^\halpha$ and $N^\halpha_E = n_E u_E^\halpha$,
the requirement that the right hand side of Eq.~\eqref{eq:marle2} vanishes imposes:
\begin{equation}
 n_E = n, \qquad u_E^\halpha = u_\halpha.
\end{equation}
Furthermore, using Eq.~\eqref{eq:SET_eq_trace} and by contracting Eq.~\eqref{eq:SET}, 
Eq.~\eqref{eq:marle1} reduces to:
\begin{equation}
 \frac{n m K_1(\zeta_E)}{K_2(\zeta_E)} = \en_E - 3P_E = \en - 3(P + \omegabar).\label{eq:betaE_def}
\end{equation}
It is important to note that $\beta_E = \zeta_E / m$, defined by Eq.~\eqref{eq:betaE_def}, does not in general 
coincide with the inverse temperature $\beta$ of the system, which is defined in terms of 
the energy density $\en$ corresponding to the stress-energy tensor $T^{\halpha\hbeta}$ 
computed from $f$. 

The simplified version of the Chapman-Enskog procedure is performed in three steps: 
first, $f$ is considered to be close to $\feq$, in which case it can be written as
\begin{equation}
 f = \feq(1 + \phi),
\end{equation}
where $\phi$ is regarded as a small number. Second, the relaxation time $\tau$ is also considered 
to be small, such that the leading constribution on the left-hand side of Eq.~\eqref{eq:boltz_cons_marle}
is given by $\feq$:
\begin{equation}
 \nabla_\mu p^\halpha e_\halpha^\mu \feq - p^{\hatt} \frac{\partial}{\partial p^\hati} 
 \left(\Gamma^\hati{}_{\halpha\hbeta} \frac{p^\halpha p^\hbeta}{p^{\hatt}} \feq\right) =
 -\frac{m}{\tau} \feq \phi.\label{eq:boltz_cons_marle_ce}
\end{equation}
In the third step, Eq.~\eqref{eq:coll_inv} is used to determine the evolution equations of the 
equilibrium quantities $n$, $u^\halpha$ and $\en_E$:
\begin{subequations}\label{eq:D_macro_eq}
\begin{align}
 D n =& -n \nabla_\hgamma u^\hgamma, \label{eq:D_n_eq}\\
 D u^\halpha =& -\frac{1}{\en_E + P_E} \Delta^{\halpha\hgamma} \nabla_\hgamma P_E,\label{eq:D_u_eq}\\
 D\en_E =& -(\en_E + P_E) \nabla_\hgamma u^\hgamma,\label{eq:D_eps_eq}
\end{align}
\end{subequations}
where
\begin{equation}
 D \equiv u^\hgamma \nabla_\hgamma
\end{equation}
is the convective derivative \cite{cercignani02,rezzolla13}. 
Combining Eqs.~\eqref{eq:D_n_eq} and \eqref{eq:D_eps_eq}, the convective derivative of the equilibrium temperature 
$T_E = \beta_E^{-1}$ can be obtained:
\begin{equation}
 DT_E = -\frac{1}{\beta_E c_{v;E}} \nabla_\hgamma u^\hgamma,\label{eq:D_T_eq}
\end{equation}
where $c_{v;E} \equiv \frac{1}{n} (\partial \en_E / \partial T_E)$ is the heat capacity, which 
has the following expression:
\begin{equation}
 c_{v,E} = \zeta_E^2 + 5 \zeta_E G_E - \zeta_E^2G_E^2 - 1,\label{eq:cv_def}
\end{equation}
where $G_E \equiv G(\zeta_E)$ is defined in Eq.~\eqref{eq:G_def}.

In the fourth step, the non-equilibrium part $\delta T^{\halpha\hbeta} \equiv T^{\halpha\hbeta} - T^{\halpha\hbeta}_E$
of the stress-energy tensor is calculated by integrating Eq.~\eqref{eq:boltz_cons_marle_ce} after a 
multiplication by $p^\halpha p^\hbeta$:
\begin{equation}
 -\frac{m}{\tau} \delta T^{\halpha \hbeta} = \nabla_\hgamma T^{\halpha\hbeta\hgamma}_E.
 \label{eq:marle_deltaT}
\end{equation}
The third order moment $T^{\halpha\hbeta\hgamma}_E$ of $\feq$ is known analytically \cite{anderson74b,cercignani02}
and has the following expression:
\begin{multline}
 T^{\halpha\hbeta\hgamma}_E = nm^2 \left[\frac{K_4(\zeta_E)}{K_2(\zeta_E)} u^\halpha u^\hbeta u^\hgamma\right.\\ 
 \left.+ \frac{G_E}{\zeta_E}(u^\halpha \eta^{\hbeta\hgamma} + u^\hbeta \eta^{\hgamma\halpha} +
 u^\hgamma \eta^{\halpha\hbeta})\right].
\end{multline}
Performing the contractions in Eqs.~\eqref{eq:SET_noneq} on Eq.~\eqref{eq:marle_deltaT} gives:
\begin{subequations}
\begin{align}
 -\frac{1}{\tau} (\en - \en_E) =& nD\left(\frac{3G_E}{\beta_E}\right) -
 2 P_E G_E \frac{Dn}{n},\label{eq:marle_deps}\\
 -\frac{1}{\tau}(P + \omegabar - P_E) =& nD\left(\frac{G_E}{\beta_E}\right) +
 \frac{2}{3} P_E G_E \nabla_\hgamma u^\hgamma,\label{eq:marle_omegabar_def}\\
 -\frac{1}{\tau} q^\hgamma =& nm\left(1 + \frac{5G_E}{m\beta_E}\right) Du^\hgamma \nonumber\\
 &+ \Delta^{\hgamma\hbeta} \nabla_\hbeta \left(\frac{\en_E + P_E}{m\beta_E}\right),\label{eq:marle_qalpha}\\
 -\frac{1}{\tau} \pi^{\halpha\hbeta} =& 2P_E G_E \nabla^{<\halpha}u^{\hbeta>}.\label{eq:marle_pi}
\end{align}
\end{subequations}

Replacing the convective derivative $Dn$ from Eq.~\eqref{eq:marle_deps} with the right hand side of 
Eq.~\eqref{eq:D_n_eq} shows that Eq.~\eqref{eq:betaE_def} indeed holds, allowing $\omegabar$ to be
cast in the form:
\begin{equation}
 \omegabar = \frac{1}{3}(\en - \en_E) - (P - P_E).\label{eq:marle_omegabar_aux}
\end{equation}
The difference $P - P_E$ can be expressed in terms of the 
difference $\en - \en_E$ by expanding $\en$ in powers of $\beta^{-1} - \beta^{-1}_E$,
and retaining only the first order term, as follows \cite{cercignani02}:
\begin{equation}
 \en - \en_E = (P - P_E) c_{v,E} + \dots.\label{eq:marle_depP}
\end{equation}
Substituting Eq.~\eqref{eq:marle_depP} in Eq.~\eqref{eq:marle_omegabar_aux} gives (to first order in 
$\beta^{-1} - \beta_E^{-1}$):
\begin{equation}
 \omegabar = \frac{c_{v,E} - 3}{3c_{v,E}} (\en - \en_E).
\end{equation}
A tedious but straightforward calculation, involving the use of Eqs.~\eqref{eq:D_n_eq} 
and \eqref{eq:D_T_eq} to eliminate the convective derivatives in Eq.~\eqref{eq:marle_deps},
yields the following expression for the coefficient of bulk viscosity:
\begin{multline}
 \eta = \frac{\tau P_E(3 - c_{v,E})}{3c_{v,E}^2} \left(20 G_E + 3\zeta_E - 3\zeta_E G_E^2\right.\\
 \left. - 2\zeta_E^2 G_E - 10\zeta_E G_E^2 + 2\zeta_E^2 G_E^3\right).\label{eq:eta}
\end{multline}

To set Eq.~\eqref{eq:marle_qalpha} in the form in Eq.~\eqref{eq:lambda_def}, the convective derivative $Du^\hgamma$
can be replaced using Eq.~\eqref{eq:D_u_eq}, while the following identities can be employed in the second term:
\begin{multline}
 \Delta^{\hgamma\hbeta} \nabla_\hbeta \left(\frac{\en_E + P_E}{m\beta_E}\right) =
 \Delta^{\hgamma\hbeta} \nabla_\hbeta \left(P_E G_E\right)\\
 = \Delta^{\hgamma\hbeta} [G_E \nabla_\hbeta P_E + \frac{1}{m} (c_{v;E} + 1)\nabla_\hbeta T_E].
\end{multline}
The coefficient of thermal conductivity can now be obtained:
\begin{equation}
 \lambda = \frac{\tau P_E}{m} (1 + c_{v;E}).\label{eq:lambda}
\end{equation}

Finally, the coefficient of shear viscosity can be read by comparing 
Eqs.~\eqref{eq:pi_def} and \eqref{eq:marle_pi}:
\begin{equation}
 \mu = \tau P_E G_E.\label{eq:mu}
\end{equation}

\begin{figure}
\begin{center}
\includegraphics[width=0.8\columnwidth]{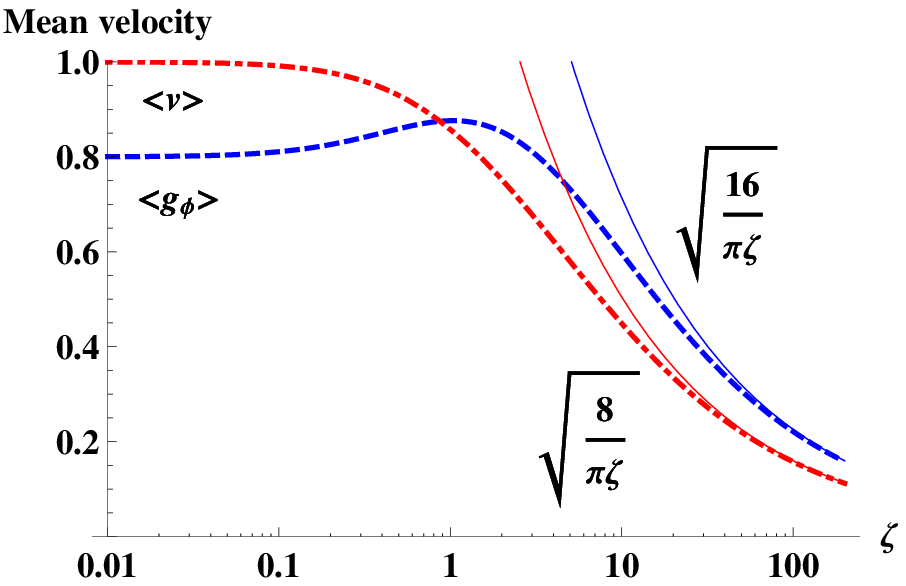}
\end{center}
\caption{Comparison between the mean of the M\"oller velocity $\braket{g_\phi}$ \eqref{eq:vmoller} 
and of the modulus of the velocity $\braket{v}$ \eqref{eq:vmean} as functions of the relativistic coldness 
$\zeta$. The limits \eqref{eq:meanv_as} at small and large $\zeta$ are confirmed: $\braket{g_\phi}$ 
goes to $\frac{4}{5}$ at small $\zeta$ and is well approximated by $\sqrt{16/\pi\zeta}$ at large $\zeta$; 
while $\braket{v}$ goes to $1$ (the speed of light) as $\zeta \rightarrow 0$, while at large $\zeta$, 
it behaves like $\sqrt{8/\pi\zeta}$.}
\label{fig:meanv}
\end{figure}

The final ingredient necessary to interpret the transport coefficients is the definition of a relaxation time.
According to Ref.~\cite{cercignani02}, the generalisation of the relaxation time to the relativistic case 
yields the following expression for $\tau$:
\begin{equation}
 \tau = \frac{1}{n\pi a^2 \mathcal{V}},\label{eq:tau}
\end{equation}
where the mean velocity $\mathcal{V}$ can be taken to represent either the average of the 
M\"oller velocity $g_\phi = \sqrt{(\bm{v} - \bm{v}_*)^2 - (\bm{v} \times \bm{v}_*)^2}$, or of 
the modulus of the velocity $\bm{v} = c\bm{p}/p^{\hatt}$:
\begin{subequations}\label{eq:meanv}
\begin{align}
 \braket{g_\phi} =& \frac{2}{\zeta_E^2 [K_1(\zeta_E)]^2} \left[
 4\zeta_E^2 {\rm Ki}_2(2\zeta_E) + 6\zeta_E {\rm Ki}_3(2\zeta_E)\right.\nonumber\\
 &+ (3 - 4\zeta_E^2) {\rm Ki}_4(2\zeta_E)
 -6\zeta_E {\rm Ki}_5(2\zeta_E)\nonumber\\
 &\left.- 3{\rm Ki}_6(2\zeta_E)\right],\label{eq:vmoller}\\
 \braket{v} =& \frac{\zeta_E}{K_1(\zeta_E)} \left[e^{-\zeta_E} \frac{1+\zeta_E}{\zeta_E^2} - 
 \Gamma(0, \zeta_E)\right],\label{eq:vmean}
\end{align}
\end{subequations}
where ${\rm Ki}_n(z)$ is the repeated integral of $K_0(z)$, defined as \cite{abramowitz72,cercignani02,olver10}:
\begin{equation}
 {\rm Ki}_n(z) = \int_0^\infty \frac{e^{-z \cosh t}}{(\cosh t)^n} dt,
\end{equation}
while $\Gamma(\nu, z)$ denotes the incomplete Gamma function \cite{abramowitz72,olver10}:
\begin{equation}
 \Gamma(\nu, z) = \int_z^\infty dt\, e^{-t} t^{\nu - 1}.
\end{equation}
Figure~\ref{fig:meanv} shows the dependency of $\braket{g_\phi}$ and $\braket{v}$ on the 
relativistic coldness $\zeta$, confirming the following limits for Eqs.~\eqref{eq:meanv}:
\begin{subequations}\label{eq:meanv_as}
\begin{align}
 \braket{g_\phi} \xrightarrow[\zeta \ll 1]{} & \frac{4}{5}, & 
 \braket{g_\phi} \xrightarrow[\zeta \gg 1]{} & \sqrt{\frac{16}{\pi\zeta}}, \label{eq:vmoller_as}\\
 \braket{v} \xrightarrow[\zeta \ll 1]{} & 1, & 
 \braket{v} \xrightarrow[\zeta \gg 1]{} & \sqrt{\frac{8}{\pi\zeta}}.\label{eq:vmean_as}
\end{align}
\end{subequations}
It is also interesting to note that the maximum value of $\braket{g_\phi}$ is attained at $\zeta_{\rm max} \simeq 1.034$,
where $\braket{g_\phi} \simeq 0.876$.

\begin{figure}
\begin{center} 
\begin{tabular}{c}
 \includegraphics[width=0.8\columnwidth]{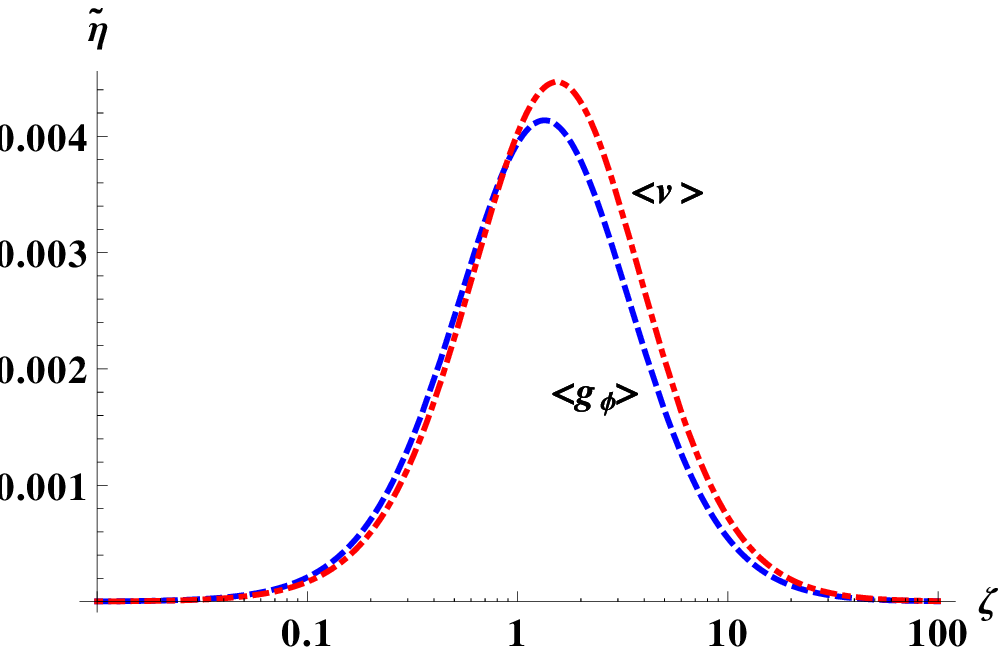} \\
 (a) \\
 \includegraphics[width=0.8\columnwidth]{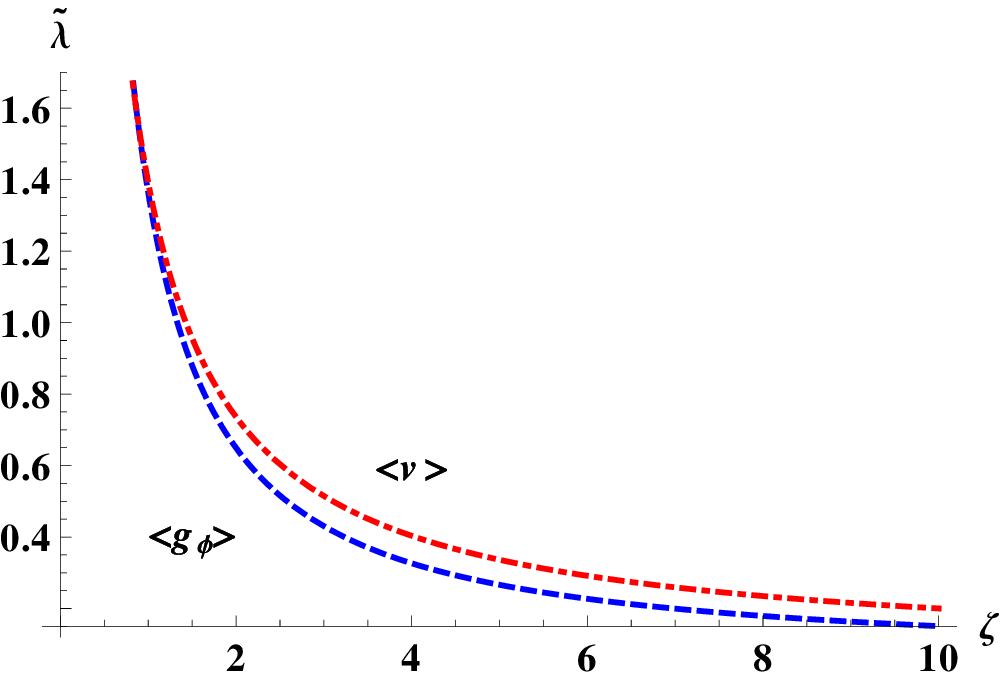} \\
 (b) \\
 \includegraphics[width=0.8\columnwidth]{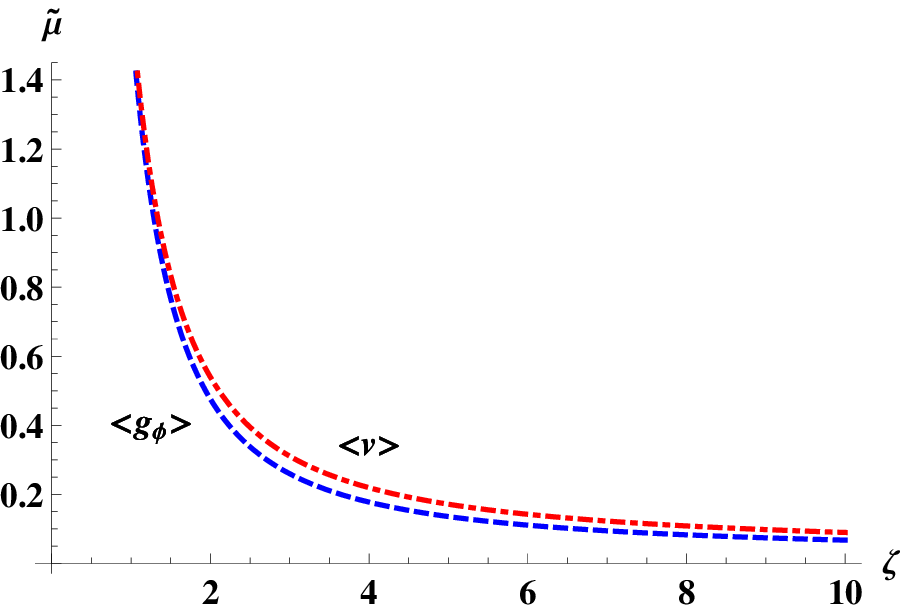} \\
 (c)
\end{tabular}
\end{center}
\caption{Plots of (a) the effective bulk viscosity $\widetilde{\eta} \equiv a^2 \eta / m$; 
(b) the thermal conductivity $\widetilde{\lambda} \equiv a^2 \lambda$; and (c) the shear 
viscosity $\widetilde{\mu} \equiv a^2\mu / m$. Each plot shows two curves, corresponding to 
the cases when the mean velocity in Eq.~\eqref{eq:tau} is taken to be the average of the M\"oller
velocity $\braket{g_\phi}$ and the average of the particle velocity $\braket{v}$, respectively.
It can be seen that, in both cases, the coefficient of shear viscosity has a maximum located 
at $\zeta_{\braket{g_\phi}} = 1.342$ and $\zeta_{\braket{v}} = 1.535$, having the values 
$\widetilde{\eta}(\zeta_{\braket{g_\phi}}) = 3.292 \times 10^{-4}$ and 
$\widetilde{\eta}(\zeta_{\braket{v}}) = 3.554 \times 10^{-4}$, respectively.}
\label{fig:tcoeff_eff}
\end{figure}

Using Eq.~\eqref{eq:tau} for the definition of $\tau$, it is convenient to introduce the following notation:
\begin{equation}
 \widetilde{\eta} = \frac{a^2 \eta}{m}, \qquad 
 \widetilde{\lambda} = a^2 \lambda, \qquad 
 \widetilde{\mu} = \frac{a^2 \mu}{m},
 \label{eq:tcoeff_eff}
\end{equation}
where the ``effective'' transport coefficients $\widetilde{\eta}$, $\widetilde{\lambda}$ 
and $\widetilde{\mu}$ only depend on $\zeta_E$ (since $P_E = n m / \zeta_E$). 
Figure~\ref{fig:tcoeff_eff} shows a comparison of Eqs.~\eqref{eq:tcoeff_eff} when the 
mean velocity is taken to be the average of the M\"oller velocity or the average of $v$, 
in terms of $\zeta$. It can be seen that, while $\widetilde{\lambda}$ and $\widetilde{\mu}$ 
decrease monotonically from infinite values at $\zeta \rightarrow 0$ to $0$ as $\zeta \rightarrow \infty$, 
the ``effective''bulk viscosity $\widetilde{\eta}$ presents a maximum value at $\zeta = \zeta_{\rm max}$,
while decreasing to $0$ as $\zeta \rightarrow 0$ or $\zeta \rightarrow \infty$.
The value of $\zeta_{\rm max}$ depends on the definition of the mean velocity, having the 
value $\zeta_{\braket{g_\phi}} = 1.342$ and $\zeta_{\braket{v}} = 1.535$, when the mean 
velocity is taken as $\braket{g_\phi}$ and $\braket{v}$, respectively. A direct comparison of the 
curves for the transport coefficients corresponding to the relaxation time constructed using 
$\braket{g_\phi}$ and $\braket{v}$ reveals that their qualitative behaviour is the same in both 
cases. Thus, for the remainder of this paper, we will only discuss the case when $\braket{v}$ 
is employed.

To conclude this section, it is worth emphasizing that the tetrad formalism has made possible the analogy
between the computation of the transport coefficients on curved spaces with respect to arbitrary coordinate 
systems and on Minkowski space in Cartesian coordinates. Moreover, the expressions \eqref{eq:eta}, \eqref{eq:lambda} 
and \eqref{eq:mu} for the coefficients of bulk viscosity, thermal conductivity and shear viscosity are 
identical to those obtained for Minkowski space \cite{cercignani02}, in agreement with Einstein's equivalence principle. 
This is not surprising, since the equations
\eqref{eq:transp_coef_def} defining the transport coefficients, as well as Eq.~\eqref{eq:marle_deltaT} describing
the non-equilibrium part of the stress-energy tensor, are written in a covariant form, reducing to 
the Minkowski expressions presented in Ref.~\cite{cercignani02} in the flat-space limit. The effect of curvature is 
however felt through the covariant derivatives in Eqs.~\eqref{eq:transp_coef_def}, which define the transport 
coefficients. 
It is worth writing down the expression for $\nabla_\hgamma u^\hgamma$ appearing in Eq.~\eqref{eq:eta_def}:
\begin{equation}
 \nabla_\hgamma u^\hgamma = e_\hgamma^\mu \partial_\mu u^\hgamma + \Gamma^\hgamma{}_{\hbeta\hgamma} u^\hbeta =
 \Gamma^\hgamma{}_{\hatt\hgamma},
\end{equation}
since, in the comoving frame, $u^\hgamma = (1,0,0,0)$. 

We end this section by noting that the expressions that we obtained for the connection coefficients 
depend on the form of the constitutive equations. In this section, we defined the transport coefficients 
using Eqs.~\eqref{eq:transp_coef_def}, which represent the covariant form of the standard definitions 
on Minkowski space \cite{rezzolla13}. While other definitions of the transport coefficients are 
possible~\cite{kremer13,kremer14}, in this paper we only consider the covariant formalism presented 
in this section.
%

\section{Rotating flows in central charts}\label{sec:sph}

In this section, we consider an application of the formalism presented in Sec.~\ref{sec:boltz} to 
the case of flows undergoing rigid rotation on spherically symmetric spacetimes. 
In Subsec.~\ref{sec:sph:vel}, the expression of the inverse temperature $\beta$ and $4$-velocity 
$u^\mu$ are found by solving the Killing equation \eqref{eq:killing}. Subsection~\ref{sec:sph:frame}
defines the comoving frame using the Lorentz boost \eqref{eq:boost} introduced in 
Subsec.~\ref{sec:boltz:comoving}. Subsection~\ref{sec:sph:eq} ends this section 
with a discussion of the form of the rigidly-rotating equilibrium states on arbitrary static
spherically-symmetric spacetimes.

\subsection{Four-velocity}\label{sec:sph:vel}

Let us consider a central chart (i.e.~static and spherically symmetric) whose metric in spherical coordinates 
$(x^{\mu})= (t,r,\theta,\varphi)$ may be written in the general form
\begin{equation}
 ds^{2} = w^{2}\left[-dt^2 + \frac{dr^2}{u^2} +
 \frac{r^2}{v^2} (d\theta^{2} + \sin^{2}\theta d\varphi^{2})\right],
 \label{eq:ds2}
\end{equation}
where $u$, $v$ and $w$ depend only on the radial coordinate $r$. The non-vanishing Christoffel symbols corresponding to 
the above metric are given below (the prime denotes differentiation with respect to $r$):
\begin{gather}
 \Gamma^t{}_{tr} = \frac{w'}{w}, \qquad \Gamma^r{}_{tt} = u^2\frac{w'}{w}, \qquad
 \Gamma^r{}_{rr} = \frac{w'}{w} - \frac{u'}{u},\nonumber\\
 \Gamma^r{}_{\theta\theta} = \frac{u^2 r^2}{v^2} \left(\frac{w'}{w} + \frac{1}{r}- \frac{v'}{v}\right),\qquad 
 \Gamma^{\theta}{}_{\varphi\varphi} = -\sin\theta\cos\theta\nonumber\\
 \Gamma^r{}_{\varphi\varphi} = -\frac{u^2 \rho^2}{v^2} \left(\frac{w'}{w} + \frac{1}{r} - \frac{v'}{v}\right), \qquad
 \Gamma^\varphi{}_{\theta\varphi} = \cot\theta,\nonumber\\
 \Gamma^\theta{}_{r\theta} = \Gamma^\varphi{}_{r\varphi} = \frac{w'}{w} + \frac{1}{r} - \frac{v'}{v}. 
\end{gather}

For the remainder of this paper, we will consider rigidly rotating flows rotating with constant 
angular velocity $\Omega$ about the $z$ axis. The only non-vanishing components of the $4$-velocity 
of such flows are $u^t$ and $u^\varphi$, which can be found once $k^\mu = (k^t, 0, 0, k^\varphi)^T$ 
is known. Substituting $(\mu, \nu) = (t, r)$ in Eq.~\eqref{eq:killing} gives:
\begin{equation}
 k_t = C_1 w^2(r),
\end{equation}
where $C_1$ is an integration constant.
Furthermore, setting $(\mu, \nu) = (r, \varphi)$ in Eq.~\eqref{eq:killing} gives:
\begin{equation}
 k_{\varphi} = \Theta(\theta) \left(\frac{w r}{v}\right)^2.
\end{equation}
The function $\Theta(\theta)$ can be determined by setting $(\mu, \nu) = (\theta, \varphi)$:
\begin{equation}
 \Theta(\theta) = C_2 \sin^2\theta,
\end{equation}
where $C_2$ is an integration constant. Let us consider the norm of $k^\mu$:
\begin{equation}
 k^2 \equiv g_{\mu\nu} k^\mu k^\nu = -C_1^2 w^2 + C_2^2 \left(\frac{w \rho}{v}\right)^2,
\end{equation}
where $\rho = r \sin\theta$ represents the distance to the $z$ axis.
Since $k^2 = -\beta^2$, it is convenient to set $C_1 = -\beta_0$ and $C_2 = \beta_0 \Omega$, such that:
\begin{subequations}
\begin{align}
 k^\mu =& \beta_0 (1, 0, 0, \Omega)^T, \label{eq:kmu}\\
 \beta \equiv \beta(r, \theta) =& \beta_0 w \sqrt{1 - \left(\frac{\rho \Omega}{v}\right)^2}.
 \label{eq:beta}
\end{align}
\end{subequations}
The velocity field $u^\mu$ can be obtained by dividing $k^\mu$ \eqref{eq:kmu} by $\beta$ \eqref{eq:beta}:
\begin{equation}
 u^\mu = \frac{\gamma}{w(r)} (1, 0, 0, \Omega)^T,\label{eq:umu}
\end{equation}
where the Lorentz factor $\gamma$ is defined as:
\begin{equation}
 \gamma = \frac{1}{\displaystyle \sqrt{1 - \left(\frac{\rho\Omega}{v}\right)^2}}.\label{eq:gamma}
\end{equation}

\subsection{Comoving frame}\label{sec:sph:frame}

In this subsection, we follow the steps in section~\ref{sec:boltz:comoving} in order to define a comoving 
tetrad for the problem of rigidly rotating flows described in the previous subsection. The first step 
is to construct a tetrad with respect to which the spacetime metric \eqref{eq:ds2} is diagonal. Such
a local frame is that of the diagonal gauge, defined as
\begin{align}
 e_{\tildet} =& \frac{1}{w} \partial_t, & 
 \omega^{\tildet} =& w dt, \nonumber\\
 e_{\tilde{r}} =& \frac{u}{w} \partial_r, & 
 \omega^{\tilde{r}} =& \frac{w}{u} dr, \nonumber\\
 e_{\tilde{\theta}} =& \frac{v}{rw} \partial_\theta, & 
 \omega^{\tilde{\theta}} =& \frac{rw}{v} d\theta, \nonumber\\
 e_{\tilde{\varphi}} =& \frac{v}{\rho w} \partial_\varphi, & 
 \omega^{\tilde{\varphi}} =& \frac{\rho w}{v} d\varphi.
 \label{eq:tetrad_tilde}
\end{align}
With respect to the above tetrad, the flow four-velocity \eqref{eq:umu} has the following components:
\begin{equation}
 u^\ta = \gamma(1, 0, 0, \frac{\rho \Omega}{v})^T,
\end{equation}
where we remind the reader that $\rho = r\sin\theta$ is the distance to the $z$ axis,
$\Omega$ is the angular velocity of the rotation, the Lorentz factor $\gamma$ is defined 
in Eq.~\eqref{eq:gamma} and $v\equiv v(r)$ is defined in Eq.~\eqref{eq:ds2}.
Substituting $u^\ta$ in Eq.~\eqref{eq:boost},
the following expression can be found for the Lorentz boost $L^\ta{}_\halpha$:
\begin{equation}
 L^\ta{}_\halpha = 
 \begin{pmatrix}
  \gamma & 0 & 0 & {\displaystyle \frac{\gamma \rho\Omega}{v}} \\
  0 & 1 & 0 & 0\\
  0 & 0 & 1 & 0\\
  {\displaystyle \frac{\gamma \rho\Omega}{v}} & 0 & 0 & \gamma
 \end{pmatrix}.
\end{equation}
The comoving frame vectors can now be calculated:
\begin{align}
 e_\hatt =& \frac{\gamma}{w} (\partial_t + \Omega \partial_\varphi), \nonumber\\
 e_\hatr =& \frac{u}{w} \partial_r, \nonumber\\
 e_\htheta =& \frac{v}{rw} \partial_\theta,\nonumber\\
 e_\hvarphi =& \frac{\gamma}{w}\left(\frac{\rho\Omega}{v}\partial_t + \frac{v}{\rho} \partial_\varphi\right).
 \label{eq:tetrad_hat}
\end{align}
while the corresponding co-frame one-forms are given by:
\begin{align}
 \omega^\hatt =& \gamma w \left(dt - \frac{\rho^2\Omega}{v^2} d\varphi\right), \nonumber\\
 \omega^\hatr =& \frac{w}{u} dr, \nonumber\\
 \omega^\htheta =& \frac{rw}{v} d\theta, \nonumber\\
 \omega^\hvarphi =& \frac{\rho \gamma w}{v} (-\Omega dt + d\varphi).
\end{align}
The expression for $L^\halpha{}_\ta$ is useful in obtaining the above co-frame one-forms:
\begin{equation}
 L^\halpha{}_\ta = 
 \begin{pmatrix}
  \gamma & 0 & 0 & -{\displaystyle \frac{\gamma \rho\Omega}{v}} \\
  0 & 1 & 0 & 0\\
  0 & 0 & 1 & 0\\
  -{\displaystyle \frac{\gamma \rho\Omega}{v}} & 0 & 0 & \gamma
 \end{pmatrix}. 
\end{equation}
It is now easy to check that the spatial components of the flow four-velocity 
vanish with respect to the comoving frame:
\begin{equation}
 u^\halpha = (1, 0, 0, 0)^T.
\end{equation}

Before ending this section, it is worth giving the metric \eqref{eq:ds2} with respect to co-rotating coordinates, 
defined as $t = t_{\rm static}$ and $\varphi = \varphi_{\rm static} - \Omega t_{\rm static}$:
\begin{equation}
 ds^2 = w^2 \left[-\gamma^{-2} dt^2 + \frac{2\rho^2\Omega}{v^2} dt d\varphi + \frac{dr^2}{u^2} + \frac{r^2}{v^2} d\Omega^2\right].
\end{equation}
Thus, the co-rotating observer sees $g_{00} \rightarrow 0$ as the Killing horizon 
(i.e.~where $k^\mu = u\beta^\mu$ becomes null) is approached:
\begin{equation}
 -g_{00} = w^2 \left(1 - \frac{\rho^2 \Omega^2}{v^2}\right) = \frac{\beta^2}{\beta_0^2} = 0.
 \label{eq:horizons}
\end{equation}
It can be seen that, on these Killing horizons, the temperature $\beta^{-1}$ tends to infinity, in agreement 
with Tolman's law \cite{tolman30,tolman30b}. In Sec.~\ref{sec:app}, we will discuss the structure of these horizons 
for the particular cases of maximally-symmetric spacetimes and of the Reissner-Nordstr\"om black holes.

\subsection{Equilibrium states}\label{sec:sph:eq}

The distribution function describing equilibrium flows of perfect (i.e.~non-viscous) fluids 
is the equilibrium distribution function, which gives rise to the following 
particle current $4$-vector $N^\mu$ and stress-energy tensor $T^{\mu\nu}$:
\begin{subequations}
\begin{align}
 N^\mu_{\rm eq} =& nu^\mu, \\
 T^{\mu\nu}_{\rm eq} =& \en u^\mu u^\nu + P \Delta^{\mu\nu},
 \label{eq:tmunu_eq}
\end{align}
\end{subequations}
where the projector $\Delta^{\mu\nu}$ on the hypersurface orthogonal to $u^\mu$ is defined as
\begin{equation}
 \Delta^{\mu\nu} = g^{\mu\nu} + u^\mu u^\nu.
\end{equation}
The energy density $\en$ and hydrostatic pressure $P$ can be obtained from $T^{\mu\nu}$ using 
the following relations:
\begin{subequations}
\begin{align}
 \en =& u_\mu u_\nu T^{\mu\nu}, \label{eq:eps}\\
 P =& \frac{1}{3} \Delta_{\mu\nu} T^{\mu\nu}.
 \label{eq:Peq}
\end{align}
\end{subequations}

Let us now consider the M-J equilibrium distribution function \eqref{eq:feq_MJ} written with respect to the 
tetrad field $\{e_{\halpha}\}$, when $p_\mu u^\mu = -p^\hatt$:
\begin{equation}
 \feq \equiv \feq(Z; \beta) = \frac{Z}{(2\pi)^3}  e^{-\beta p^\hatt},
\end{equation}
where we have taken a vanishing chemical potential. 
With respect to this tetrad, $N^{\halpha}$ and $T^{\halpha\hbeta}$ in Eq.~\eqref{eq:tmunu_eq}
take the following form:
\begin{subequations}\label{eq:macro_eq}
\begin{align}
 N^{\halpha} =& \int \frac{d^3p}{p^{\hatt}} \feq\, p^{\halpha} = 
 (n, 0, 0, 0)^T,\\
 T^{\halpha\hbeta} =& \int \frac{d^3p}{p^{\hatt}} \feq\, p^{\halpha}p^{\hbeta} = 
 {\rm diag}(\en, P, P, P).
\end{align}
\end{subequations}
The integrals in the above equations can be performed analytically
in terms of modified Bessel functions \cite{cercignani02,ambrus15}:
\begin{subequations}\label{eq:hydro_MJ}
\begin{align}
 n_{\rm M-J} =& \frac{Z}{\pi^2 \beta^3} \wK_2(m \beta),\label{eq:hydro_MJ_n}\\
 \en_{\rm M-J} =& \frac{3Z}{\pi^2 \beta^4} \left[\wK_2(m \beta)
 + \frac{(m\beta)^2}{6} \wK_1(m \beta)\right],\label{eq:hydro_MJ_e}\\
 P_{\rm M-J} =& \frac{Z}{\pi^2 \beta^4} \wK_2(m \beta),\label{eq:hydro_MJ_P}
\end{align}
\end{subequations}
where $\beta = \beta_0 w \sqrt{1 - (\rho \Omega / v)^2}$ and
\begin{equation}
 \wK_n(m\beta) \equiv \frac{(m \beta)^n}{2^{n-1} (n-1)!} K_n(m \beta)
 \label{eq:widek_def}
\end{equation}
reduces to unity in the massless limit (i.e.~$m\rightarrow 0$) for all positive integers 
$n > 0$ \cite{ambrus15}. Eqs.~\eqref{eq:hydro_MJ_n} and \eqref{eq:hydro_MJ_P} confirm that the 
hydrostatic pressure is related to the particle number density $n$ and local inverse temperature $\beta$ 
through \cite{cercignani02}:
\begin{equation}
 P = n\beta^{-1}.\label{eq:Pn}
\end{equation}
Setting the mass to $0$ in Eqs.~\eqref{eq:hydro_MJ} yields:
\begin{subequations}\label{eq:hydro_MJ_m0}
\begin{align}
 n_{\rm M-J} =& \frac{Z}{\pi^2 \beta^3},\\
 \en_{\rm M-J} =& \frac{3Z}{\pi^2 \beta^4},\\
 P_{\rm M-J} =& \frac{Z}{\pi^2 \beta^4}.
\end{align}
\end{subequations}
While in this paper we only considered gas particles obeying Maxwell-J\"uttner statistics, the above 
results can readily be extended to Bose-Einstein and Fermi-Dirac statistics, as described in 
Refs.~\cite{florkowski15,ambrus15}. 

\begin{figure}
\begin{center}
\begin{tabular}{c}
 \includegraphics[width=0.75\columnwidth]{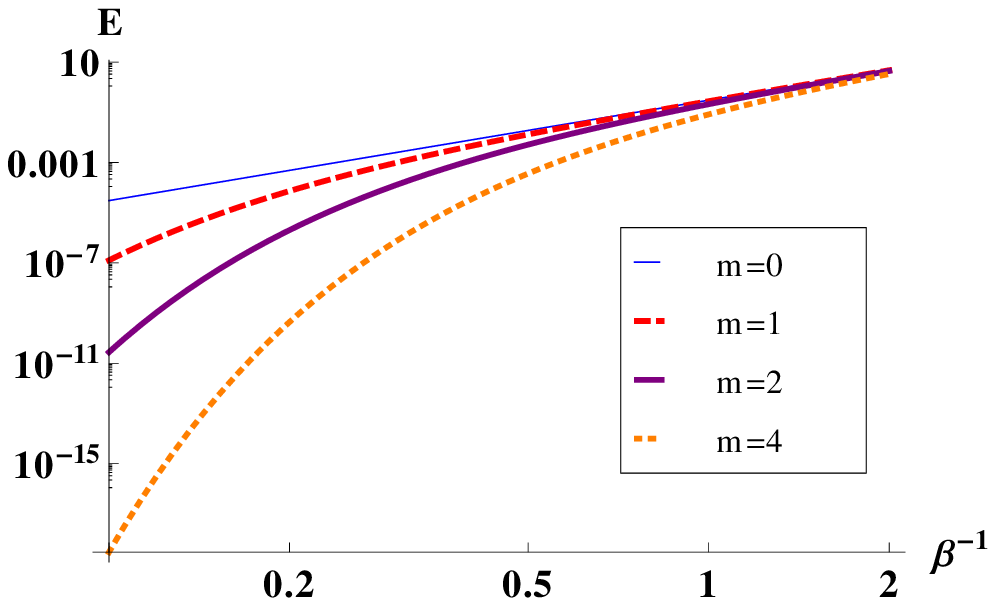}\\
 \includegraphics[width=0.75\columnwidth]{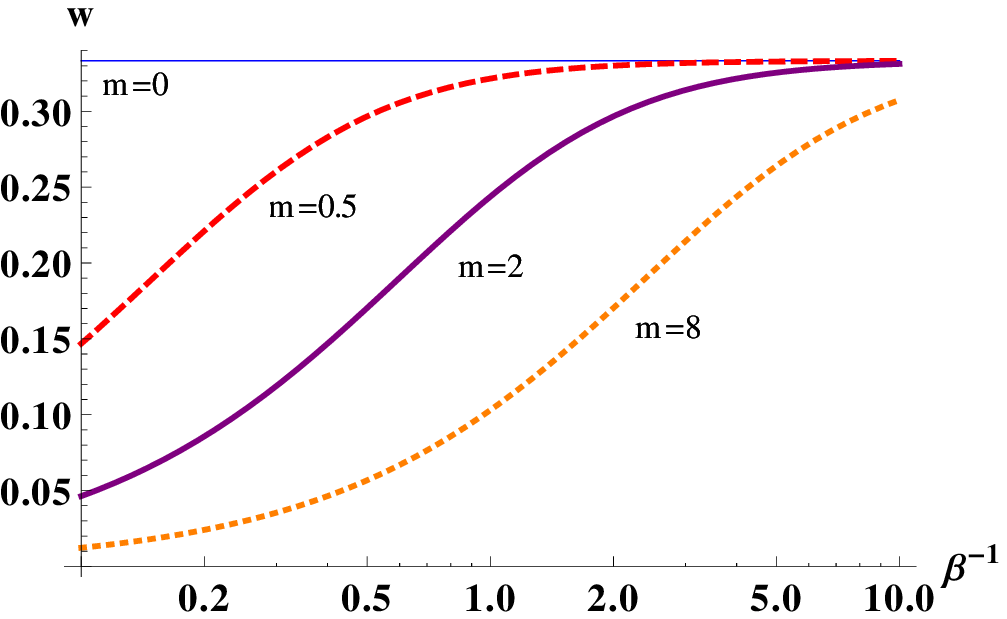}
\end{tabular}
 \caption{The dependence of the energy density $E$ (top) and equation of state $w = P/E$ (bottom)
 on the inverse temperature $\beta^{-1}$,
 for various values of the mass $m$. The expressions for $E$ and $P$ can be found in 
 Eqs.~\eqref{eq:hydro_MJ_e} and \eqref{eq:hydro_MJ_P}, respectively.
}
 \label{fig:enw}
\end{center}
\end{figure}

Since the modified Bessel functions in the expressions of $n$, $\en$ and $P$ in Eqs.~\eqref{eq:hydro_MJ}
decrease monotonically as their argument increases, it can be seen that these quantities also decrease 
monotonically with the increase of $m$ or $\beta$. 
The plots in Fig.~\ref{fig:enw} show the dependence of the energy density $E$ \eqref{eq:hydro_MJ_e} and 
equation of state $w = P/E$ with respect to the temperature $\beta^{-1}$ for various values of the mass,
confirming the monotonic behaviour of these functions as the temperature is increased.

\section{Applications}\label{sec:app}

In this section, we consider the properties of the particle flux four-vector $N^\halpha$ and stress-energy tensor 
$T^{\halpha\hbeta}$, as well as of the effective transport coefficients $\widetilde{\eta}$, $\widetilde{\lambda}$ and 
$\widetilde{\nu}$ defined in Eqs.~\eqref{eq:tcoeff_eff}.  
Since $n$, $\en$, $P$, $\widetilde{\lambda}$ and $\widetilde{\mu}$ are 
monotonic functions of $\beta$, their properties can be inferred directly from the behaviour of $\beta$. 
Thus, in this section, only the properties of $\beta$ will be presented. Since $\widetilde{\eta}$ is 
non-monotonic in $\zeta = m\beta$, its properties will also be discussed.

The analysis of $\beta$ will be focused on the structure of the Killing horizons seen by co-rotating observers, 
as described by Eq.~\eqref{eq:horizons}. Furthermore, the regimes where $\widetilde{\eta}$ is monotonic, or where 
it exhhibits regions of local extrema will be discussed. For simplicity, in this section, we only consider 
the relaxation time \eqref{eq:tau} constructed using $\braket{v}$ \eqref{eq:vmean}, since the results 
obtained using $\braket{g_\phi}$ are qualitatively similar.


According to Eq.~\eqref{eq:horizons}, the temperature measured by co-rotating observers 
diverges on the Killing horizons associated with the Killing vector in Eq.~\eqref{eq:kmu}.
In the case when the metric functions $w$ and $v$, defined in Eq.~\eqref{eq:ds2}, are non-zero 
and well defined everywhere in the spacetime, such surfaces represent speed of light surfaces
(i.e.~co-rotating observers travel at the speed of light):
\begin{equation}
 1 - \left(\frac{\rho \Omega}{v}\right)^2 = 0.
\end{equation}
The second class refers to horizons which occur in spaces where $w$ and $v$ can vanish for some choice of the spacetime 
coordinates. In the absence of rotation, they coincide with the familiar event or 
cosmological horizons, for the cases of black holes or of the de Sitter expanding universe, respectively. 

Two classes of spherically-symmetric spacetimes will be considerd in what follows: 
the maximally symmetric spacetimes (e.g.~the Minkowski, de Sitter and anti-de Sitter spacetimes) 
will be discussed in Subsec.~\ref{sec:app:msym}, while Reissner-Nordstr\"om spacetimes
will be the subject of Subsec.~\ref{sec:app:rn}.

\subsection{Maximally symmetric spaces}\label{sec:app:msym}
\begin{figure}
\begin{center}
\begin{tabular}{c}
 \includegraphics[width=0.75\columnwidth]{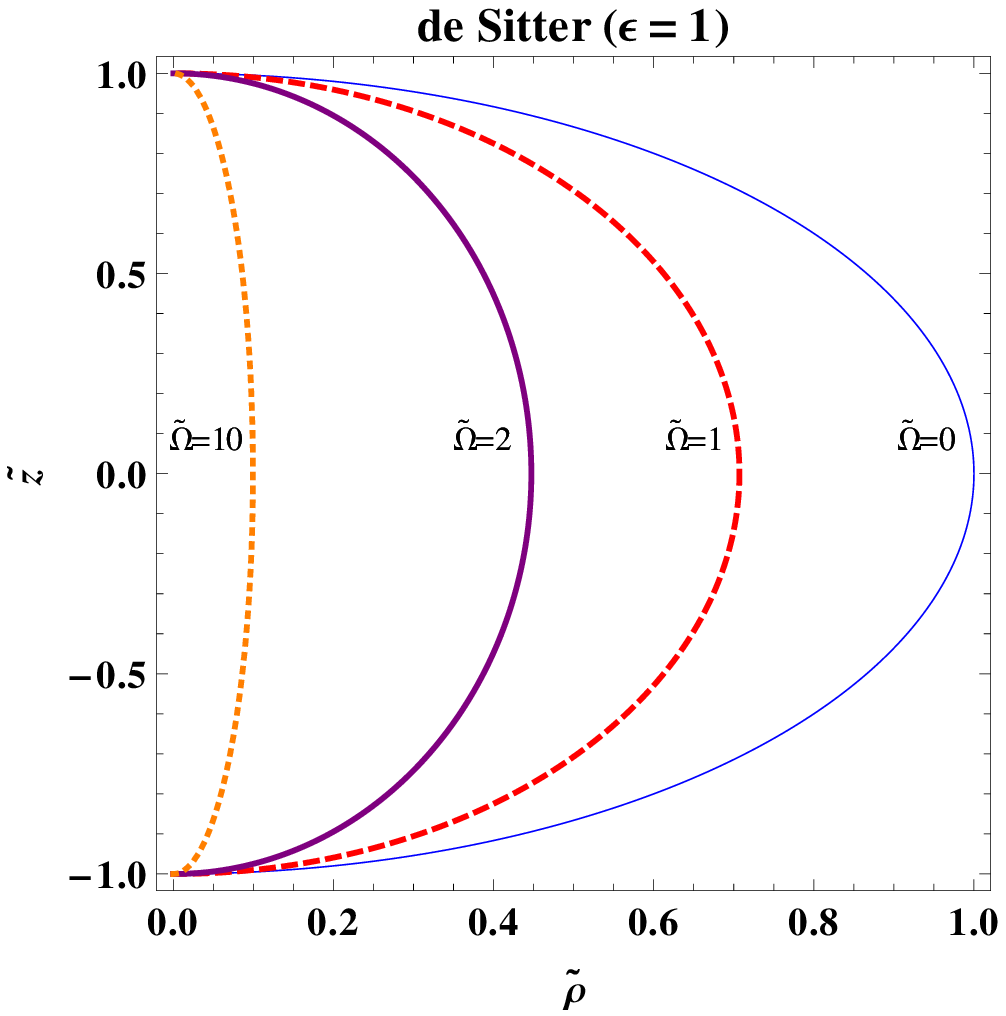}\\
 (a) \\
 \includegraphics[width=0.75\columnwidth]{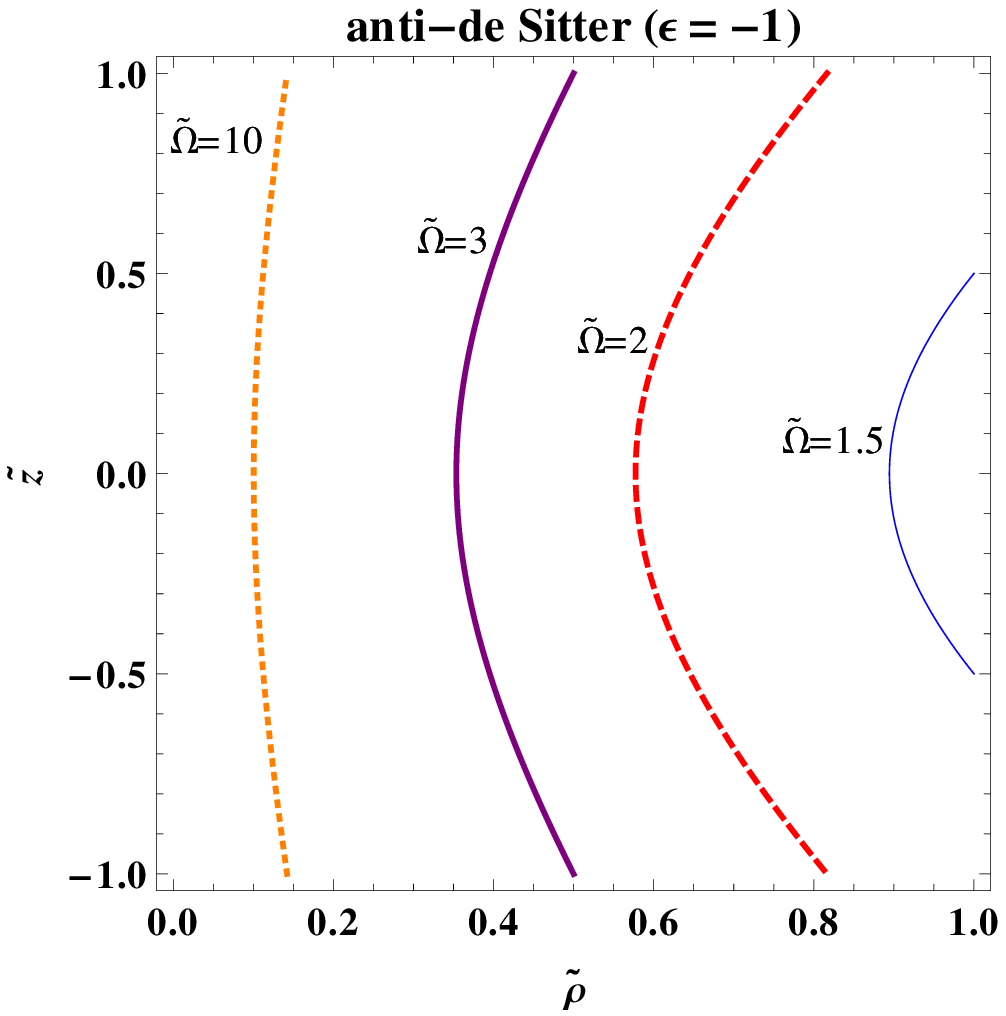}\\
 (b)
\end{tabular}
 \caption{The SOL structure of (a) dS and (b) AdS. 
 The vertical axis represents the coordinate $\widetilde{z} \equiv \omega r \cos\theta$
 along the rotation axis, while the horizontal axis represents the distance 
 $\widetilde{\rho} \equiv \omega r \sin\theta$ from the rotation axis.
 In the dS case, the rightmost line corresponding to $\widetilde{\Omega} = 0$ 
 represents the cosmological horizon, located at $\widetilde{r} = \pi / 2$.}
 \label{fig:ms_sols}
\end{center}
\end{figure}
\begin{figure}
\begin{center}
\begin{tabular}{c}
 \includegraphics[width=0.75\columnwidth]{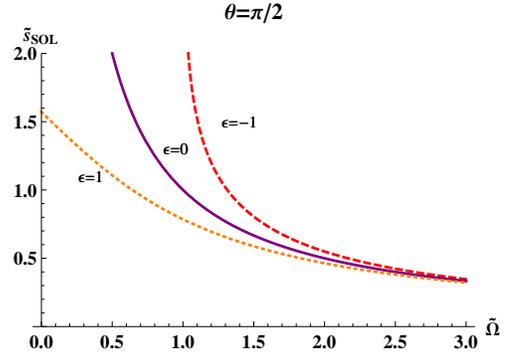}
\end{tabular}
 \caption{The proper radial distance $\widetilde{s}_{\rm SOL}$ \eqref{eq:ms_ssol} between the origin and the SOL 
 in the equatorial plane ($\sin\theta = 1$) with respect to $\widetilde{\Omega} = \Omega / \omega$. 
 The bottom, middle and top lines correspond to the cases $\epsilon = 1$ (dS), $\epsilon = 0$ (Minkowski)
 and $\epsilon = -1$ (AdS). In the case of Minkowski spacetime, we adopt the convention 
 $\widetilde{s} = s = \Omega^{-1}$ and $\widetilde{\Omega} = \Omega$ 
 (i.e.~the parameter $\omega$ is immaterial in this case).
}
 \label{fig:ms_comp}
\end{center}
\end{figure}
\begin{figure}
\begin{center}
\begin{tabular}{c}
 \includegraphics[width=0.75\columnwidth]{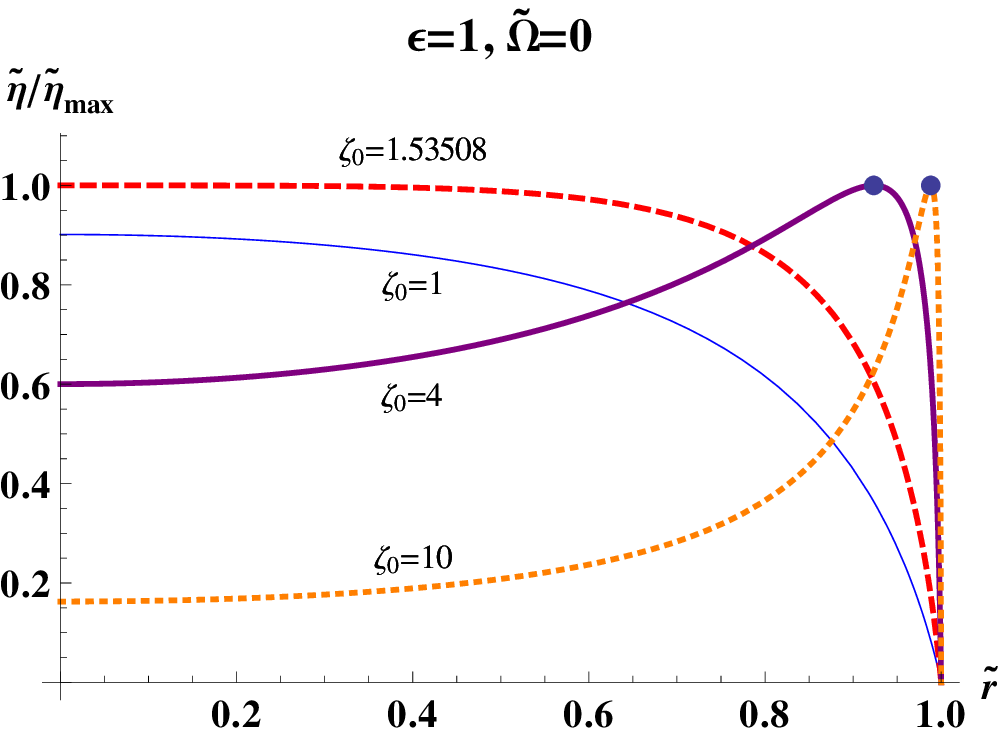}\\
 (a)\\
 \includegraphics[width=0.75\columnwidth]{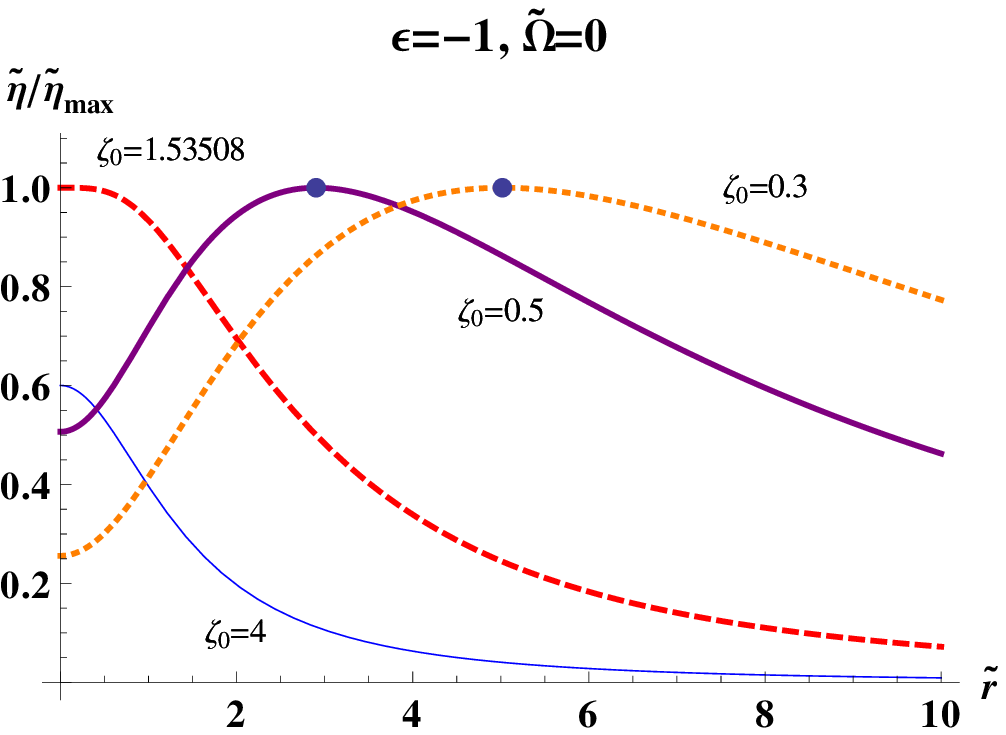} \\
 (b)
\end{tabular}
 \caption{Dependence of the effective coefficient of bulk viscosity $\widetilde{\eta}$ \eqref{eq:eta} 
 divided by its maximum value $\widetilde{\eta}_{\rm max}$ on $\widetilde{r}$ on (a) dS and (b) AdS 
 spacetimes in the absence of rotation ($\widetilde{\Omega} = 0$).
 Each curve corresponds to a different value of $\zeta_0 = m\beta_0$, where 
 $\beta_0 \equiv \beta(r = 0)$ is the inverse temperature at the 
 coordinate origin. The local maxima exhibited by $\widetilde{\eta}$ (highlighted by circular points) 
 appear only when $\zeta_0$ is large (dS) or small (adS), its locations being given 
 by Eqs.~\eqref{eq:ms_rmax_dS} and \eqref{eq:ms_rmax_adS} for the 
 dS and adS spaces, respectively.
 }
 \label{fig:ms_eta_zetas}
\end{center}
\end{figure}
\begin{figure*}
\begin{center}
\begin{tabular}{cc}
 \includegraphics[width=0.75\columnwidth]{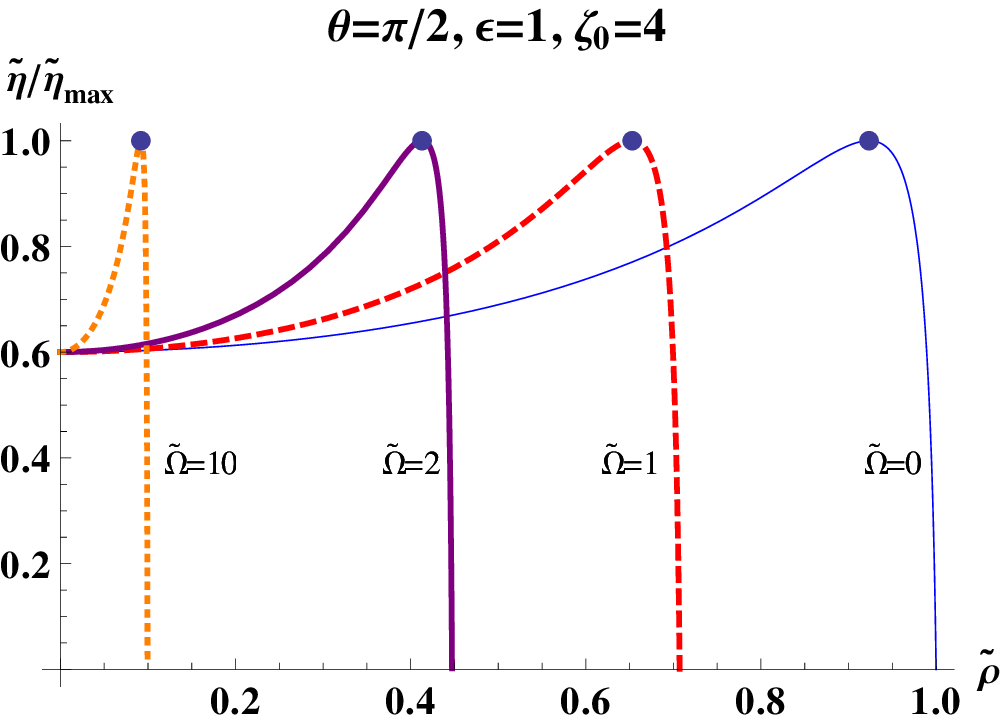} \hspace{0.05\columnwidth} &
 \hspace{0.05\columnwidth} \includegraphics[width=0.75\columnwidth]{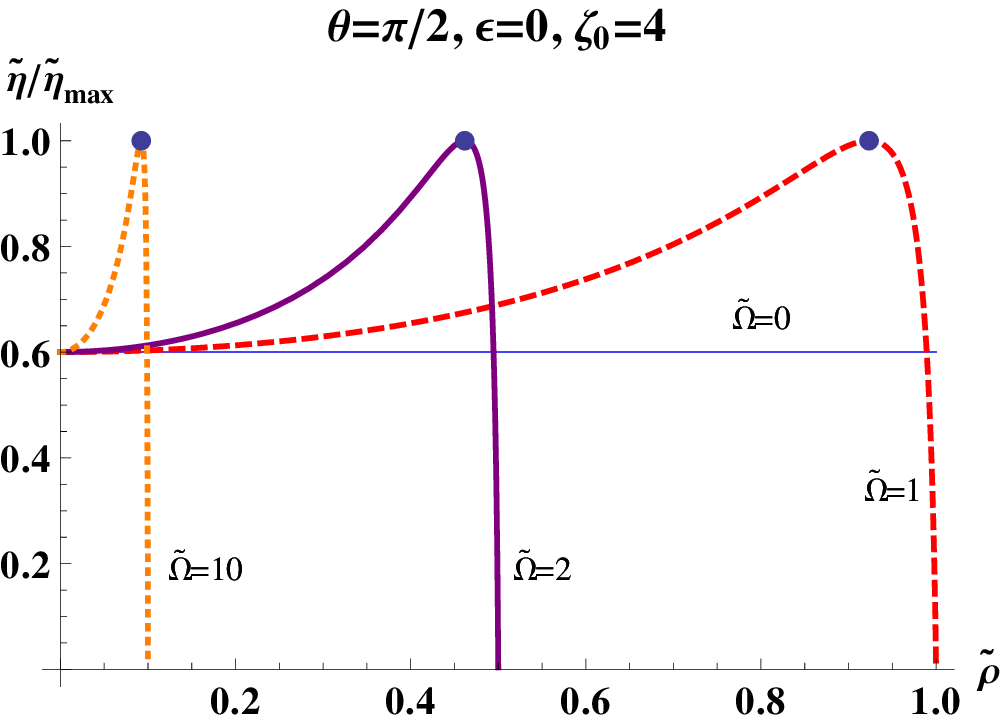}\\
 (a) &
 (b)\\
 \includegraphics[width=0.75\columnwidth]{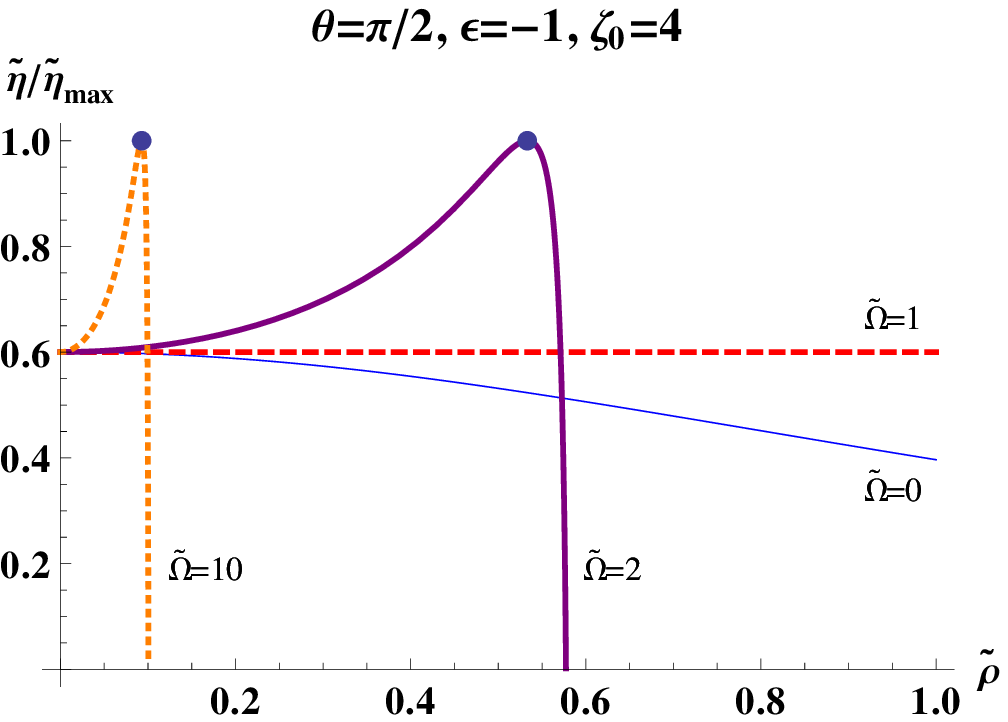} \hspace{0.05\columnwidth} &
 \hspace{0.05\columnwidth} \includegraphics[width=0.75\columnwidth]{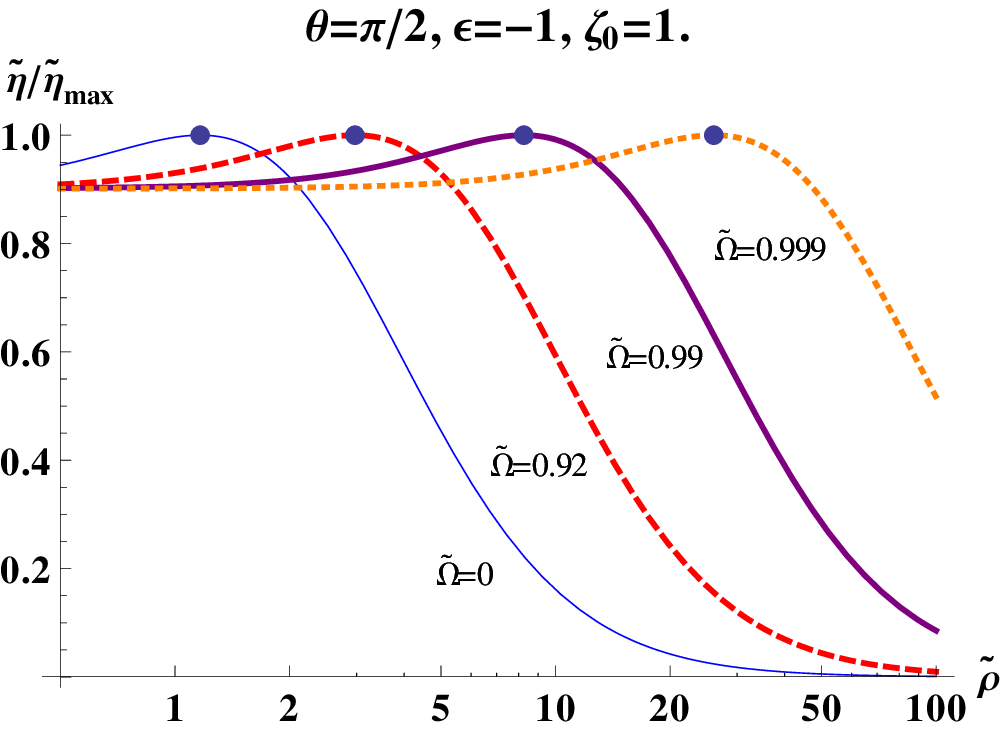}\\
 (c) &
 (d)
\end{tabular}
 \caption{Dependence of the effective coefficient of bulk viscosity $\widetilde{\eta}$ \eqref{eq:eta} 
 in the equatorial plane ($\theta = \pi/2$)
 divided by its maximum value $\widetilde{\eta}_{\rm max}$ on (a) dS, (b) Minkowski and (c)-(d) AdS spacetimes.
 Each curve corresponds to a different value of $\widetilde{\Omega}$, while the parameter $\zeta_0 = 4$ for 
 (a)-(c) and $\zeta_0 = 1$ for (d). It should be noted that when $\widetilde{\Omega} = 1$, $\widetilde{\eta}$ 
 is constant in the equatorial plane of AdS space, cf.~\eqref{eq:ads_betaconst}.
 }
 \label{fig:ms_eta_omegas}
\end{center}
\end{figure*}
The maximally-symmetric spaces which make the subject of the present subsection 
represent vacuum solutions of the Einstein equations in the presence of a cosmological 
constant equal to:
\begin{equation}
 \Lambda = 3\epsilon \omega^2,
\end{equation}
where $\epsilon=0$, $1$ and $-1$ for the Minkowski, de Sitter and anti-de Sitter spacetimes, 
respectively. The notation $\omega$ refers to the Hubble constant for de Sitter space and 
to the inverse radius of curvature for anti-de Sitter space. For completeness, we also 
give the corresponding Ricci scalar:
\begin{equation}
 R = 12\epsilon\omega^2.
\end{equation}
The line element can be written as \cite{pascu12}:
\begin{equation}
 ds^2 = -(1 - \epsilon \omega^2 r^2) dt^2 + \frac{dr^2}{1 - \epsilon \omega^2 r^2} + 
 r^2 d\Omega^2,\label{eq:ds2_msym}
\end{equation}
where the radial coordinate $r$ has the range $[0, \infty)$ on AdS and 
$r \in [0, \omega^{-1})$ on dS. On dS spacetime, the surface $r = \omega^{-1}$ represents 
the cosmological horizon.
Eq.~\eqref{eq:ds2_msym} can be put in the form of the generic line element in Eq.~\eqref{eq:ds2}
by making the following identifications:
\begin{equation}  
w=v=\sqrt{1-\epsilon \omega^2 r^2}, \qquad u=w^2,
\end{equation} 
Using Eq.~\eqref{eq:beta}, the following expression can be obtained for the inverse temperature $\beta$:
\begin{equation}
\beta=\beta_0\sqrt{1-(\epsilon \omega^2 +\Omega^2 \sin^2\theta)r^2},
\label{eq:beta_msym}
\end{equation}
where $\beta_0$ represents the inverse temperature at the origin $r = 0$.
Setting $\Omega = 0$ in Eq.~\eqref{eq:beta_msym} shows that, in the absence of rotations,
the local temperature $\beta^{-1}$ remains constant (Minkowski case), decreases to $0$ 
as $r \rightarrow \infty$ (AdS case) or increases to infinity as the cosmological 
horizon $r = \omega^{-1}$ is approached (dS case). 


When $\Omega \neq 0$, the rotation induces an SOL where $\beta$ \eqref{eq:beta_msym}
vanishes, such that:
\begin{equation}
 1 - \epsilon \widetilde{r}^2 - \widetilde{\rho}^2 \widetilde{\Omega}^2 = 0,
 \label{eq:msym_SOL}
\end{equation}
where the notation 
\begin{equation}
 \widetilde{r} = \omega r, \qquad 
 \widetilde{\rho} = \omega r \sin\theta, \qquad 
 \widetilde{\Omega} = \frac{\Omega}{\omega}.
\end{equation}
was introduced for convenience. 
In the case of the Minkowski spacetime ($\epsilon = 0$), the SOL is located where
\begin{equation}
 \rho \Omega= 1.
\end{equation}
Rearranging Eq.~\eqref{eq:msym_SOL} to
\begin{equation}
 \widetilde{\rho}^2 \widetilde{\Omega}^2 = 1 - \epsilon \widetilde{r}^2
\end{equation}
shows that the repulsive nature of a positive cosmological constant ($\epsilon = 1$), 
occuring in the case of the dS space, induces a further centrifugal effect, pulling the SOL inwards 
with increasing $r$. In the AdS case ($\epsilon = -1$), the attractive nature of a negative cosmological 
constant $\Lambda = -3\omega^2$ can play the role of a centripetal force, thus diminishing the effect
of rotation. 

The position $\widetilde{r}_{\rm SOL}$ of the SOL can be found from Eq.~\eqref{eq:msym_SOL}:
\begin{equation}
 \widetilde{r}_{\rm SOL} = \frac{1}{\sqrt{\widetilde{\Omega}^2 \sin^2\theta + \epsilon}}.
\end{equation}
On dS, the SOL always forms inside the cosmological horizon, being located at 
\begin{equation}
 \widetilde{r}_{\rm SOL} = \frac{1}{\sqrt{\widetilde{\Omega}^2 \sin^2\theta + 1}} \qquad (\text{de Sitter}).
\end{equation}
It can be seen that the SOL always touches the cosmological horizon on the rotation axis
(i.e.~$\theta = 0$), where $\widetilde{r}_{\rm SOL} = 1$. This behaviour is illustrated in 
Fig.~\ref{fig:ms_sols}(a).

The situation on AdS is quite different: as shown in Ref.~\cite{nicolaevici01},
compact manifolds do not exhibit superluminal velocities unless the rotation parameter 
is sufficiently large. In this case, no SOL forms if $\widetilde{\Omega} < 1$
and the temperature $\beta^{-1}$ remains finite throughout the spacetime. 
For $\widetilde{\Omega} \ge 1$, the location of the SOL is given by:
\begin{equation}
 \widetilde{r}_{\rm SOL} = \frac{1}{\sqrt{\widetilde{\Omega}^2 \sin^2\theta - 1}} \qquad (\text{anti-de Sitter}),
\end{equation}
where $\theta$ is constrained such that $\sin\theta \ge \widetilde{\Omega}^{-1}$, as shown in 
Fig.~\ref{fig:ms_sols}(b). Furthermore, Eq.~\eqref{eq:msym_SOL} implies that for a fixed value of 
$\widetilde{\Omega}$, the value of $\beta$ (and indeed of all quantities derived from it, such as 
$n$, $E$, $P$, $\widetilde{\eta}$, $\widetilde{\mu}$ and $\widetilde{\lambda}$)
is constant on the cone having its apex at the origin, for which 
\begin{equation}
 \sin\theta = \widetilde{\Omega}^{-1}. \label{eq:ads_betaconst}
\end{equation}
In particular, setting $\widetilde{\Omega} = 1$ implies that $\beta$ is constant throughout the 
equatorial plane. 

A more geometric assessment of the location of the SOL is the proper radial distance $\widetilde{s} = \omega s$ 
from the origin to the SOL, which can be written as follows:
\begin{align}
 \widetilde{s} =& \omega \int_0^{\widetilde{r}_{\rm SOL}} d\widetilde{r} \sqrt{g_{rr}}\nonumber\\
 =& 
 \begin{cases}
  {\rm arcsin}\, \frac{1}{\displaystyle \sqrt{\widetilde{\Omega}^2 \sin^2\theta + 1}}, & ({\rm dS})\\
  \widetilde{\Omega}^{-1}, & ({\rm Minkowski})\\
  {\rm arcsinh}\, \frac{1}{\displaystyle \sqrt{\widetilde{\Omega}^2 \sin^2\theta -1}}. & ({\rm AdS})
 \end{cases}
 \label{eq:ms_ssol}
\end{align}
Figure~\ref{fig:ms_comp} shows that, for fixed $\widetilde{\Omega}$, the distance from 
the SOL to the rotation axis in the equatorial plane is larger in the AdS and smaller in the dS cases
with respect to the same distance in Minkowski space.

The plots in Fig.~\ref{fig:ms_eta_zetas} show the dependence of $\widetilde{\eta}$ on $\overline{r}$ 
for various values of the relativistic coldness $\zeta_0 = m \beta_0 \equiv m \beta(r = 0)$ measured at 
the origin when $\widetilde{\Omega} = 0$ for the cases of (a) the dS and (b) the adS spaces. 
On dS space, $\beta$ decreases monotonically from the maximum value $\beta_0$ at the origin towards $0$ on the 
cosmological horizon (where $\widetilde{r} = 1$). For all $\zeta \le \zeta_{\rm max} \simeq 1.53508$, 
Fig.~\ref{fig:tcoeff_eff}(a) implies that $\widetilde{\eta}$ also decreases monotonically, since in this regime,
$\widetilde{\eta}$ shows no local extrema. However, for all $\zeta_0 > \zeta_{\rm max}$, $\widetilde{\eta}$ increases 
up to the maximum value $\widetilde{\eta}_{\rm max} \simeq 3.554 \times 10^{-4}$, attained when:
\begin{equation}
 \widetilde{r}_{\rm max} = \sqrt{1 - \left(\frac{\zeta_{\rm max}}{\zeta_0}\right)^2},\label{eq:ms_rmax_dS}
\end{equation}
as can be seen in Fig.~\ref{fig:ms_eta_zetas}(a).
On adS space, $\zeta$ increases monotonically from $\zeta_0$ at the origin to infinity 
as $\widetilde{r} \rightarrow \infty$. Figure~\ref{fig:ms_eta_zetas}(b) shows that
$\widetilde{\eta}$ also decreases monotonically to $0$ as $\widetilde{r} \rightarrow \infty$ 
for all $\zeta_0 \ge \zeta_{\rm max}$, while in the case when 
$\zeta_0 < \zeta_{\rm max}$, $\widetilde{\eta}$ attains the maximum 
value $\widetilde{\eta}_{\rm max}$ when
\begin{equation}
 \widetilde{r}_{\rm max} = \sqrt{\left(\frac{\zeta_{\rm max}}{\zeta_0}\right)^2 - 1}.\label{eq:ms_rmax_adS}
\end{equation}

At non-vanishing values of $\widetilde{\Omega}$,
$\widetilde{\eta}$ attains the maximum value $\widetilde{\eta}_{\rm max}$ at
\begin{equation}
 \widetilde{r} = \left[\frac{1 - (\zeta_{\rm max} / \zeta_0)^2}{\widetilde{\Omega}^2 + \epsilon}\right]^{1/2}.
\end{equation}
For the dS space, Fig.~\ref{fig:ms_eta_omegas}(a) shows that increasing the value of $\widetilde{\Omega}$ 
decreases the distance to the horizon, while the location of the maximum also decreases according to:
\begin{equation}
 \left.\widetilde{r}_{\rm max}\right\rfloor_{\rm dS} = 
 \frac{1}{\sqrt{1 + \widetilde{\Omega}^2}}\sqrt{1 - \left(\frac{\zeta_{\rm max}}{\zeta_0}\right)^2}.
\end{equation}
Figure~\ref{fig:ms_eta_omegas}(b) shows that, on Minkowski space, $\widetilde{\eta}$ is constant throughout 
the spacetime in the absence of rotation, while for non-vanishing values of $\widetilde{\Omega}$, it attains 
a maximum for all $\zeta_0 \ge \zeta_{\rm max}$ located at:
\begin{equation}
 \left.\widetilde{r}_{\rm max}\right\rfloor_{\rm Mink} = 
 \frac{1}{\widetilde{\Omega}}\sqrt{1 - \left(\frac{\zeta_{\rm max}}{\zeta_0}\right)^2}.
\end{equation}
On AdS space, three regimes can be distinguished.
When $\widetilde{\Omega} > 1$, an SOL forms and the characteristics of $\widetilde{\eta}$ 
are similar to the case when dS space is considered. When $\widetilde{\Omega} = 1$,
$\widetilde{\eta}$ is constant throughout the equatorial plane, as implied by Eq.~\eqref{eq:ads_betaconst}.
These two regimes can be clearly seen in Fig.~\ref{fig:ms_eta_omegas}(c).
Finally, when $\widetilde{\Omega} < 1$, 
no SOL forms and $\widetilde{\eta}$ only attains a maximum when $\zeta_0 < \zeta_{\rm max}$, 
as shown in Fig.~\ref{fig:ms_eta_omegas}(d).
The position of this maximum increases as $\widetilde{\Omega}$ increases. 
\subsection{Reissner-Nordstr\"om metric}\label{sec:app:rn}
\begin{figure*}
\begin{center}
\begin{tabular}{cc}
 \includegraphics[width=0.9\columnwidth]{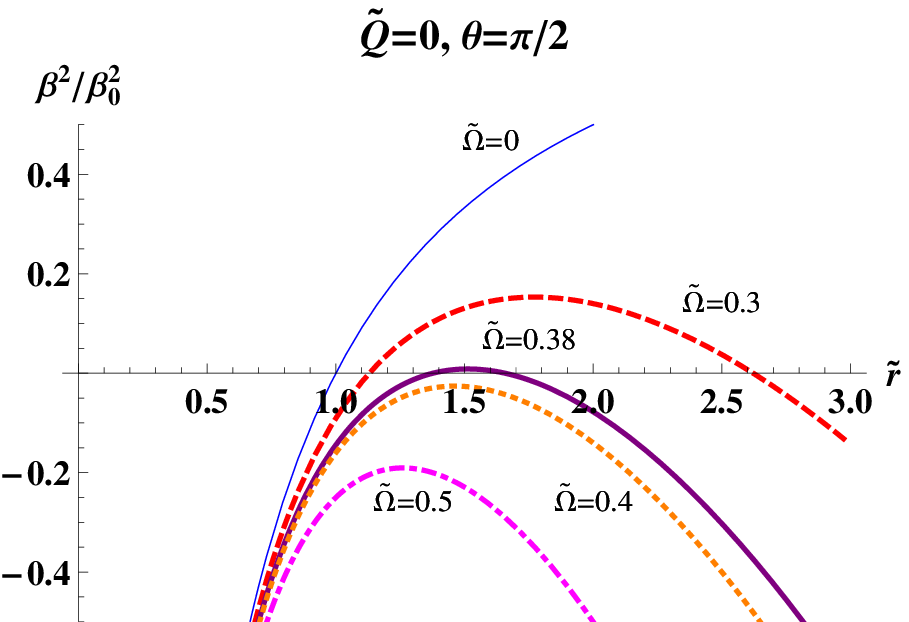} &
 \includegraphics[width=0.9\columnwidth]{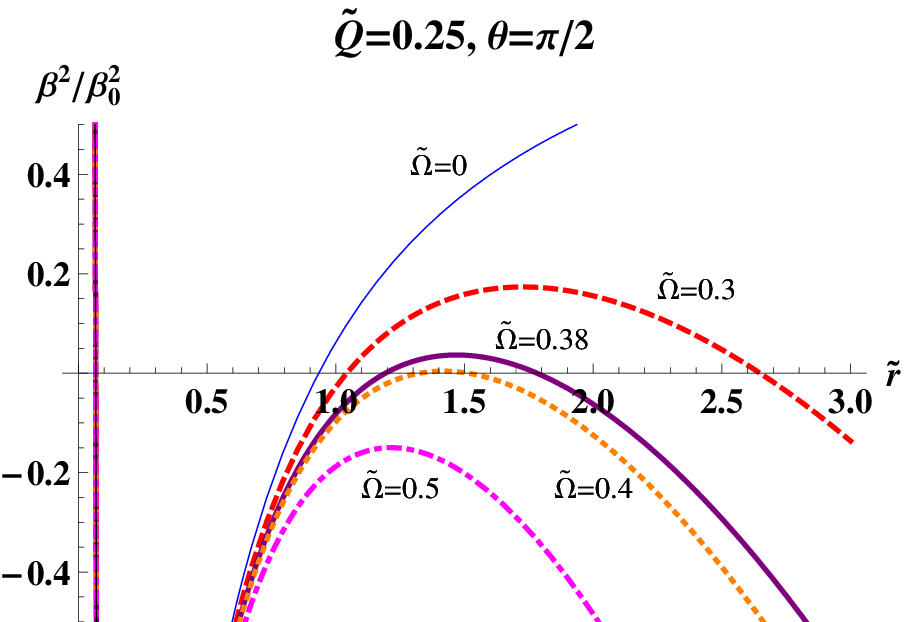}\\
 (a) & (b) \\
 \includegraphics[width=0.9\columnwidth]{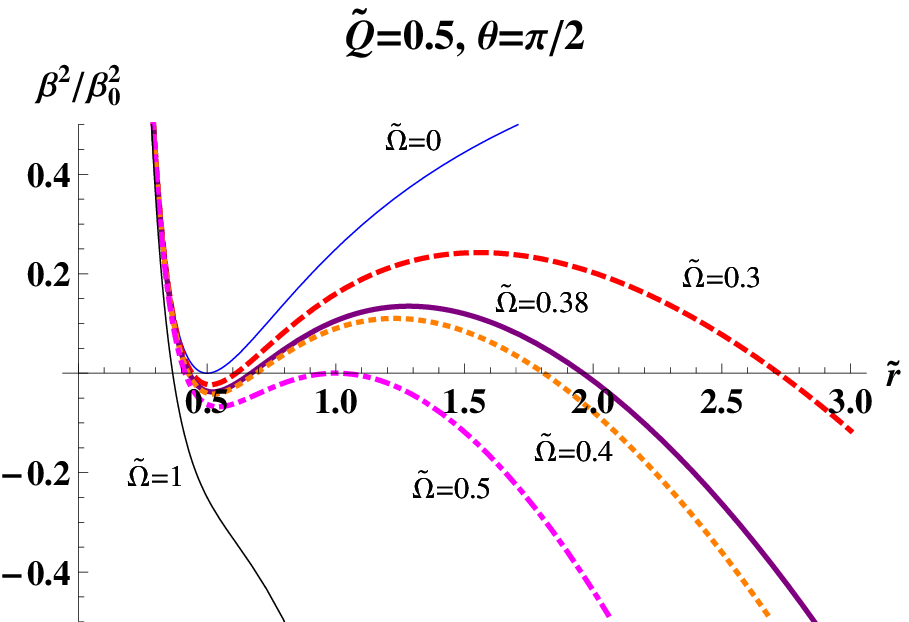} &
 \includegraphics[width=0.9\columnwidth]{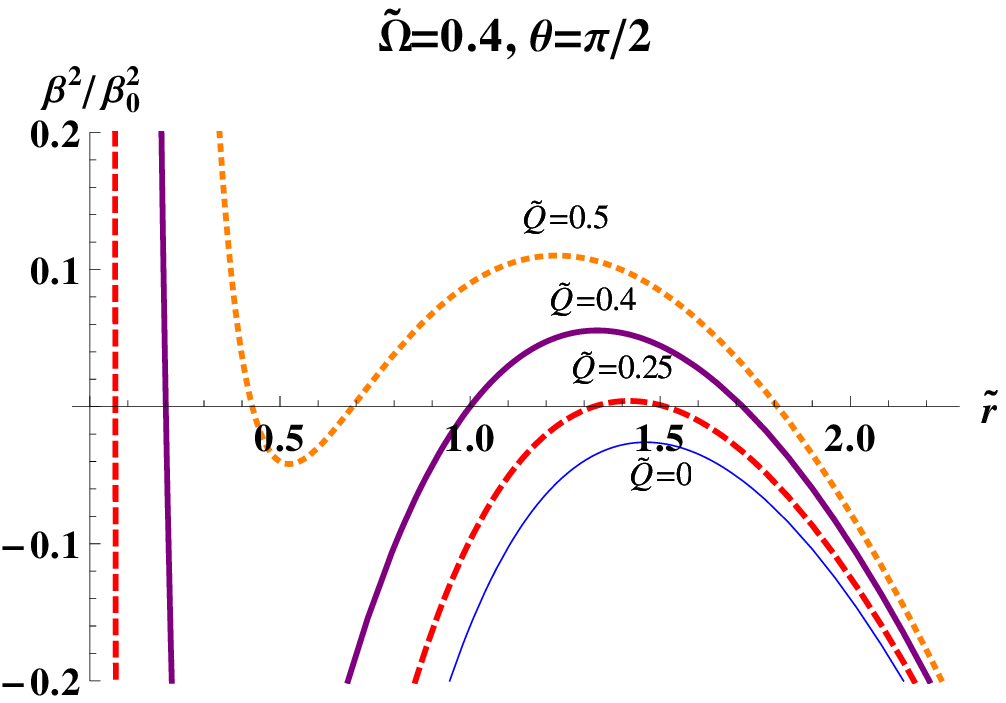} \\
 (c) & (d)
\end{tabular}
 \caption{The dependence of $\beta^2 / \beta_0^2$ on $\widetilde{r} = r/2M$ in the equatorial 
 plane $\sin\theta = 1$ for (a)-(c) $\widetilde{Q}$ fixed at (a) $0$ (Schwarzschild space),
 (b) $0.25$ and (c) $0.5$ (extremal Reissner-Nordstr\"om space), for various values of 
 $\widetilde{\Omega} = 2M \Omega$. In (d), $\widetilde{\Omega}$ is fixed at $0.4$ and the 
 charge is varied from $\widetilde{Q} = 0$ to $\widetilde{Q} = 0.5$.
 The regions where $\beta^2 > 0$ represent ``allowed'' regions (where the temperature is 
 finite and well defined, while the points where $\beta = 0$ represent horizons.}
\label{fig:rn_bsq}
\end{center}
\end{figure*}
\begin{figure*}
\begin{center}
\begin{tabular}{cc}
 \includegraphics[width=0.9\columnwidth]{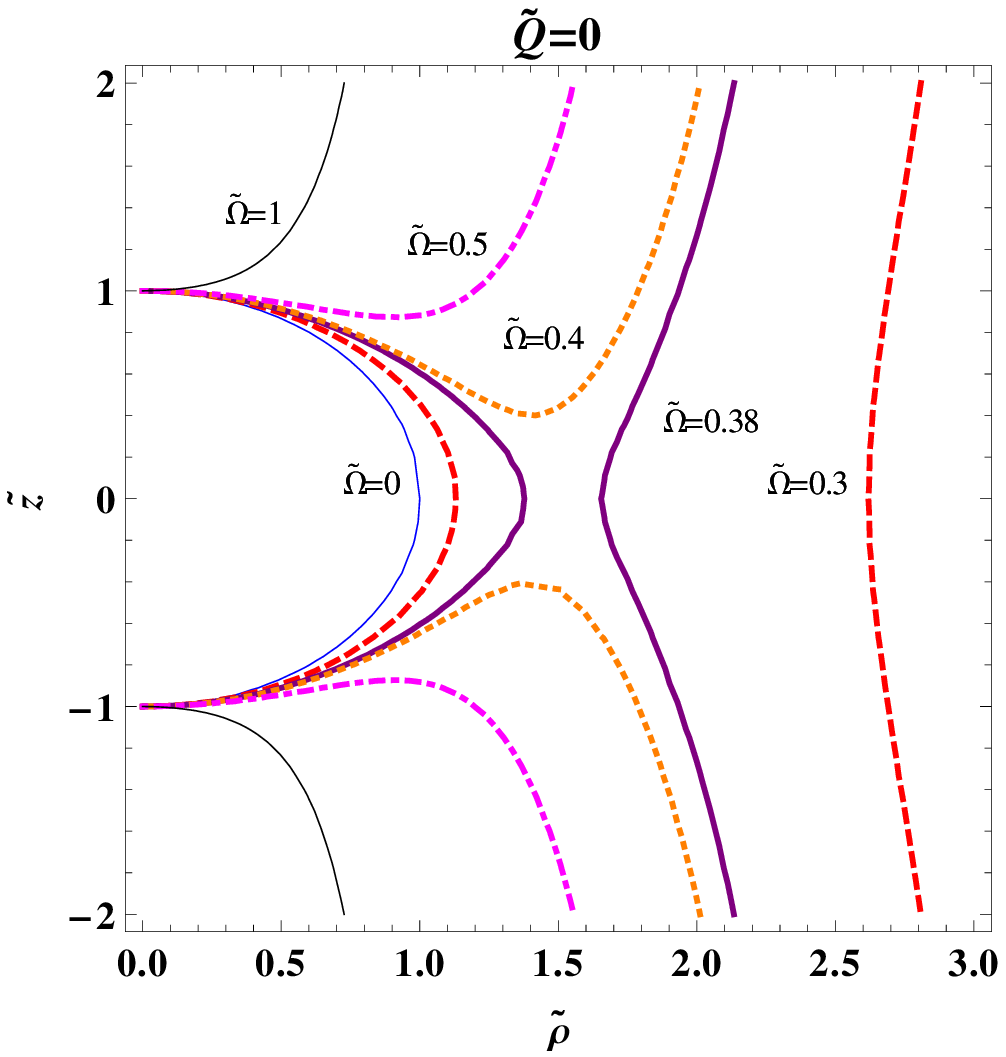} &
 \includegraphics[width=0.9\columnwidth]{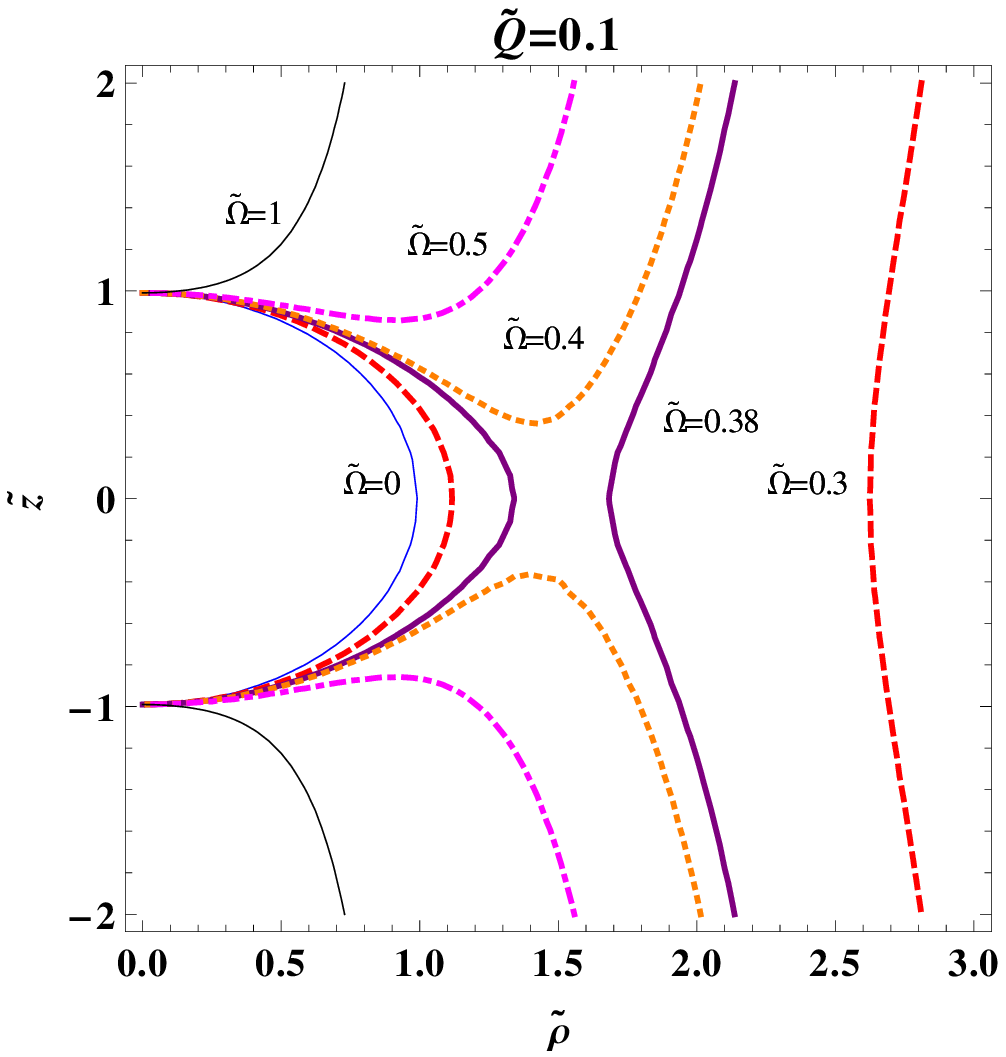}\\
 \hspace{22pt} (a) & \hspace{22pt} (b) \\
 \includegraphics[width=0.9\columnwidth]{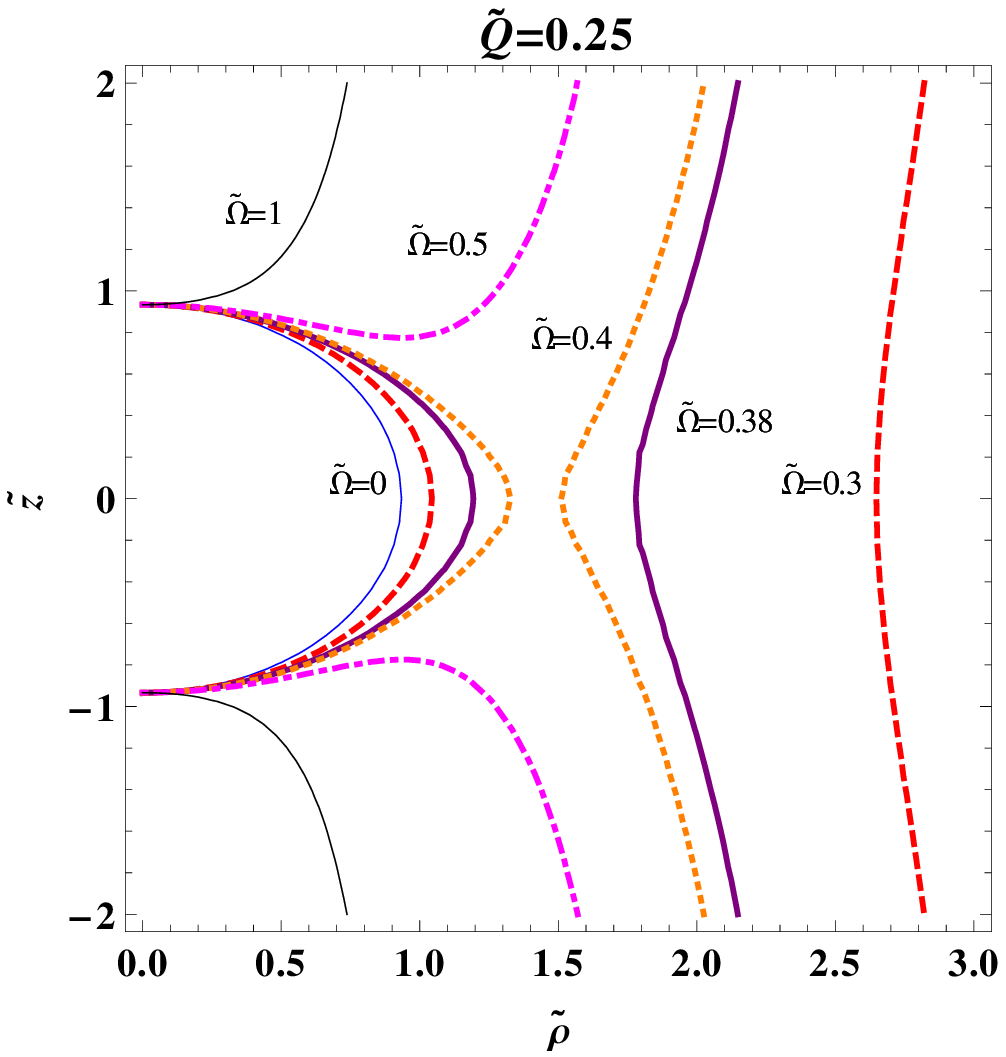} &
 \includegraphics[width=0.9\columnwidth]{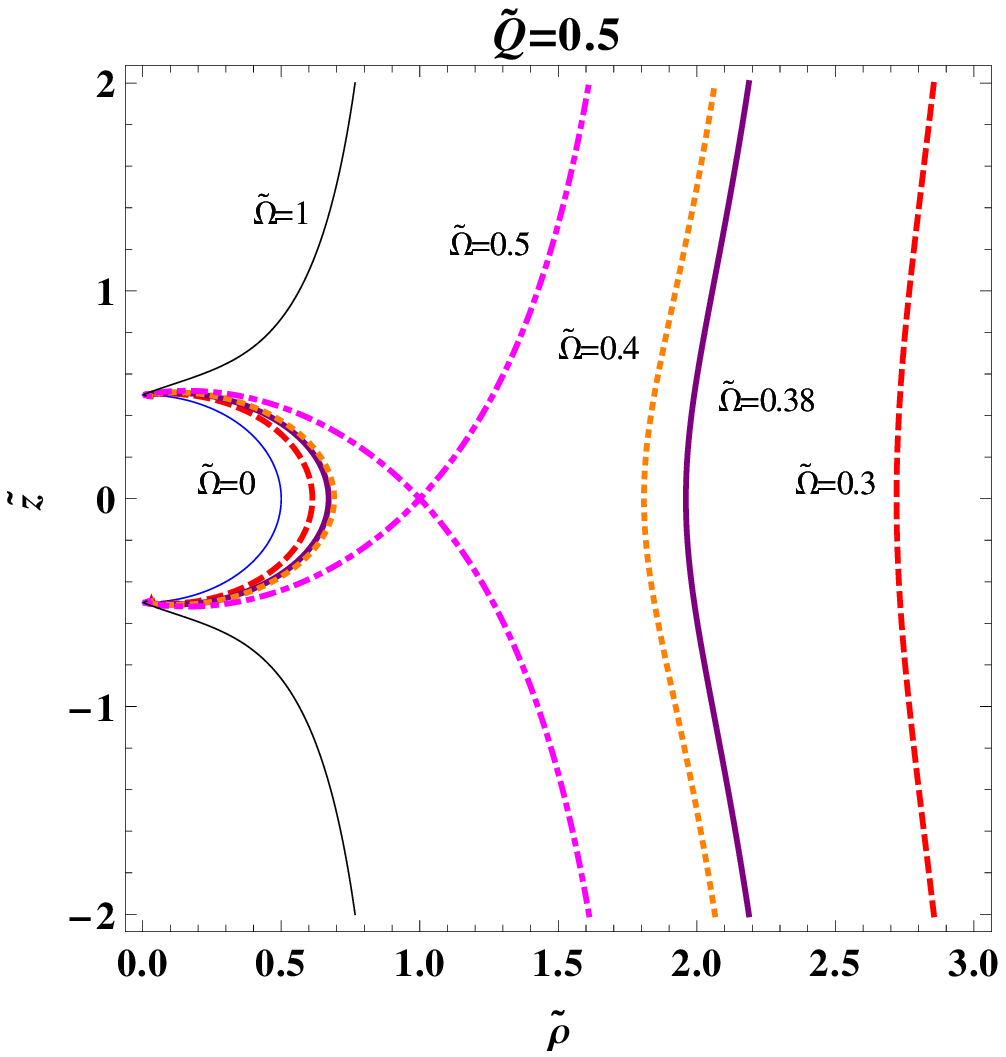}\\
 \hspace{22pt} (c) & \hspace{22pt} (d)
\end{tabular}
 \caption{The SOL structure of the Reissner-Nordstr\"om spacetime for four values of 
 $\widetilde{Q} = Q / 2M$. The vertical axis represents the coordinate $\widetilde{z} \equiv z / 2M$
 along the rotation axis, while the horizontal axis represents the distance 
 $\widetilde{\rho} = r \sin\theta / 2M$ from the rotation axis. The contours represent the surfaces where 
 $\beta = 0$, i.e.~either the black hole horizon (only the outer horizons are shown) or the rotation horizon 
 (i.e.~where the SOL induced by the rotation forms). The case (a) shows the horizon structure 
 for the Schwarzschild space ($Q = 0$), while the case (d) represents an extremal Reissner-Nordstr\"om 
 black hole ($Q = M$).}
\label{fig:rn_sol}
\end{center}
\end{figure*}
\begin{figure*}
\begin{center}
\begin{tabular}{cc}
 \includegraphics[width=0.9\columnwidth]{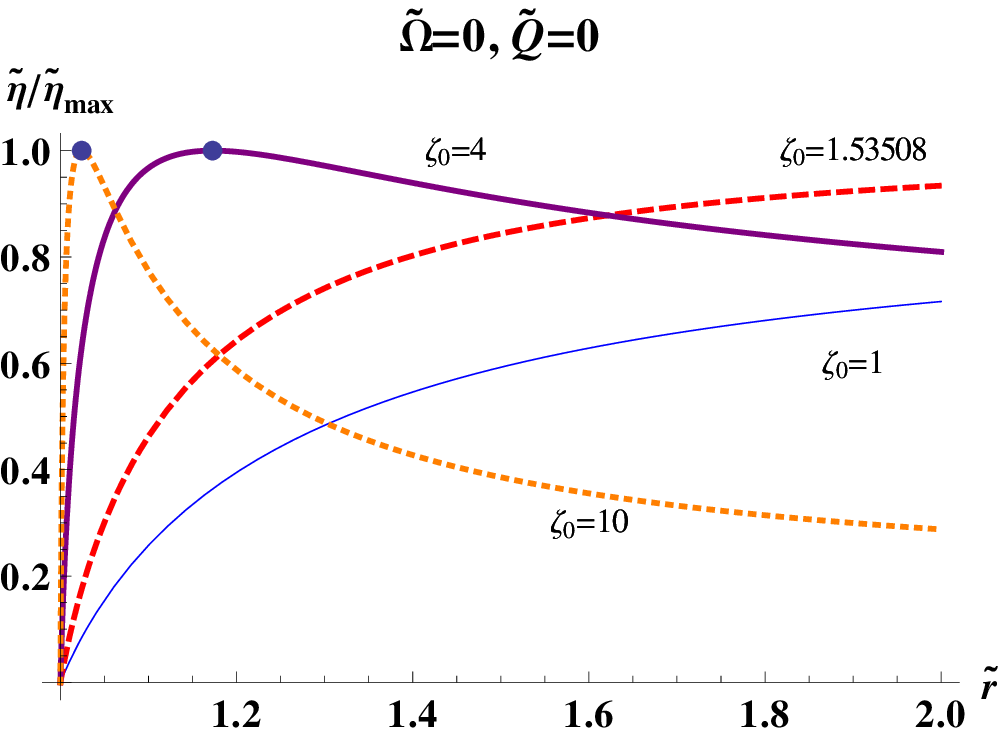} &
 \includegraphics[width=0.9\columnwidth]{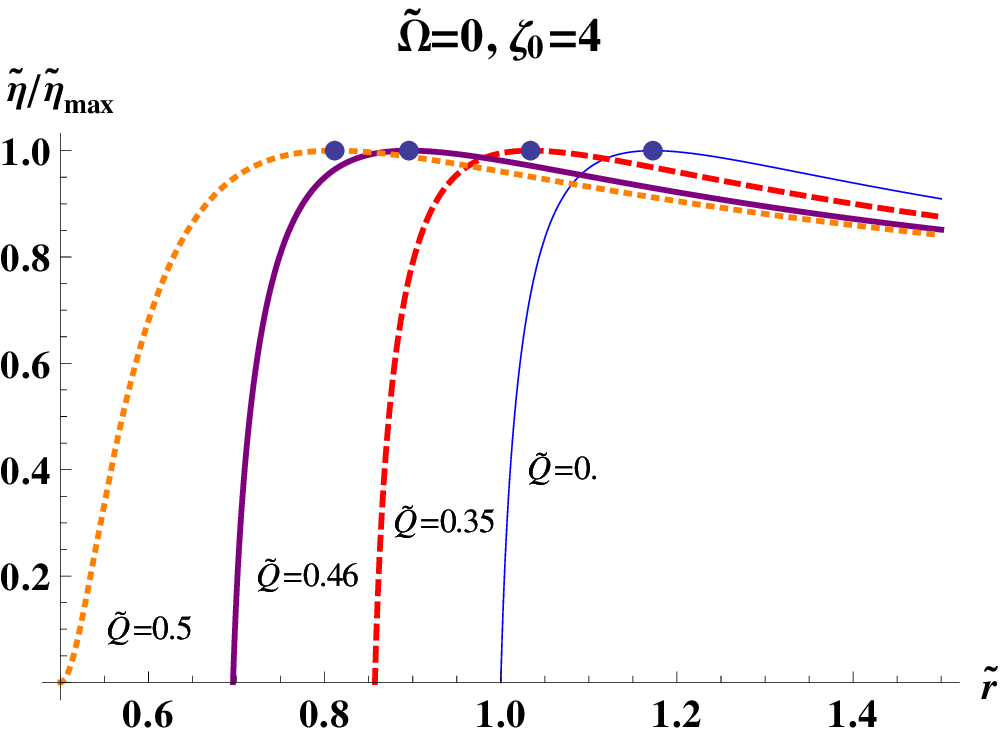}\\
 \hspace{22pt} (a) & \hspace{22pt} (b) \\
 \includegraphics[width=0.9\columnwidth]{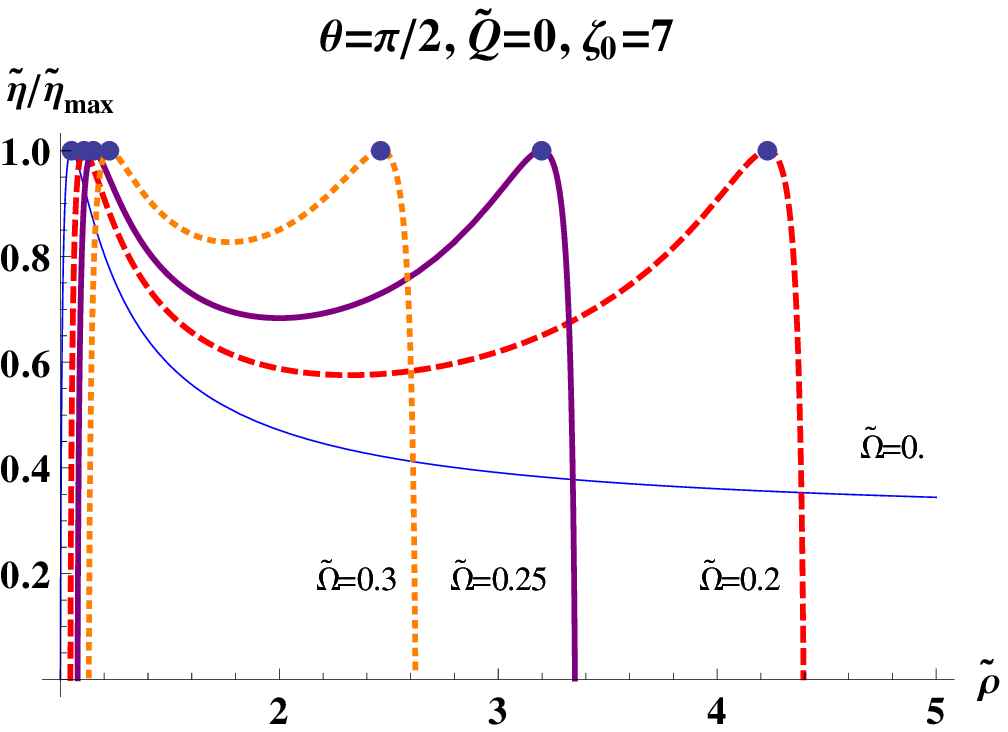} &
 \includegraphics[width=0.9\columnwidth]{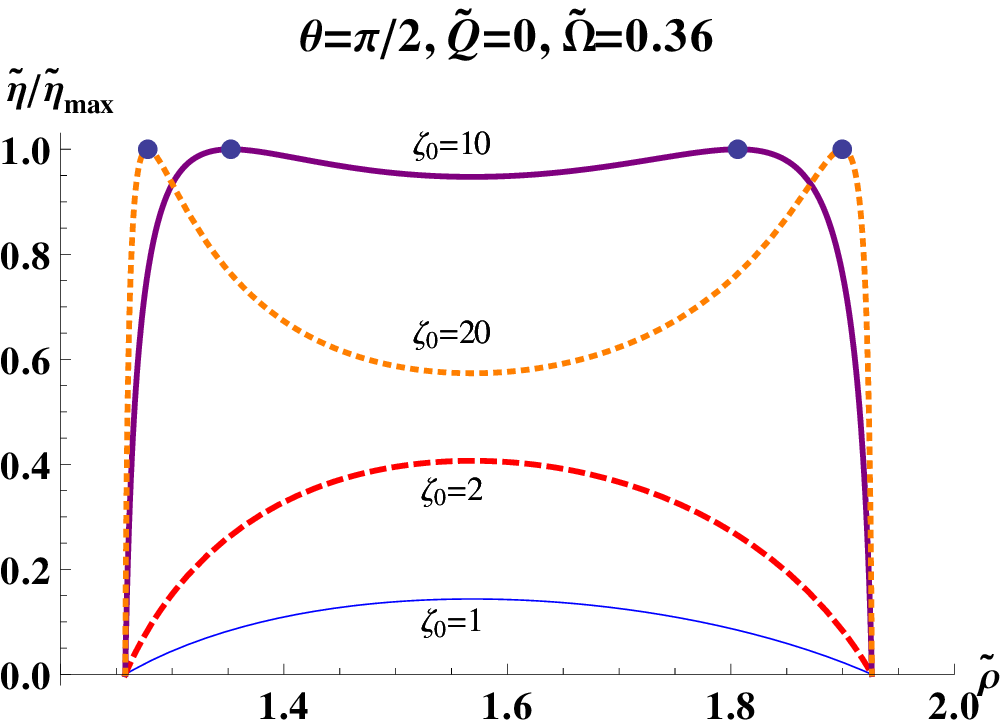}\\
 \hspace{22pt} (c) & \hspace{22pt} (d)
\end{tabular}
 \caption{Dependence of the effective coefficient of bulk viscosity $\widetilde{\eta}$ \eqref{eq:eta} 
 in the equatorial plane ($\theta = \pi/2$)
 divided by its maximum value $\widetilde{\eta}_{\rm max}$ on the Reissner-Nordstr\"om spacetime, 
 in the following cases:
 (a) $\widetilde{Q} = 0$ and $\widetilde{\Omega} = 0$ for various values of $\zeta_0$;
 (b) $\widetilde{\Omega} = 0$ and $\zeta_0 = 4$ for various values of $\widetilde{Q}$;
 (c) $\widetilde{Q} = 0$ and $\zeta_0 = 7$ for various values of $\widetilde{\Omega}$;
 (d) $\widetilde{Q} = 0$ and $\widetilde{\Omega} = 0.36$ for various values of $\zeta_0$.}
\label{fig:rn_eta}
\end{center}
\end{figure*}
The line element of the Reissner-Nordstr\"om metric is given by: 
\begin{equation}
 ds^2 = -\left(1 - \frac{2M}{r} + \frac{Q^2}{r^2} \right) dt^2 + \frac{dr^2}{1 - \frac{2M}{r} + \frac{Q^2}{r^2}} 
 + r^2 d\Omega^2,\label{eq:ds2_rn}
\end{equation}
describing the gravitational field of a black hole of mass $M$ and charge $Q$.
Comparing the above equation with Eq.~\eqref{eq:ds2} gives the following expressions 
for the metric functions:
\begin{equation}
 w = v = \sqrt{1 - \frac{2M}{r} + \frac{Q^2}{r^2}}, \qquad 
 u = 1 - \frac{2M}{r} + \frac{Q^2}{r^2},
\end{equation}
such that the inverse temperature $\beta$ \eqref{eq:beta} becomes:
\begin{equation}
 \beta = \beta_0 \sqrt{1 - \frac{2M}{r} + \frac{Q^2}{r^2} - \rho^2\Omega^2},
 \label{eq:beta_rn}
\end{equation}
where $\beta_0$ is the inverse temperature at infinity on the rotation axis (i.e.~$z = r\cos\theta \rightarrow \pm \infty$ and 
$\rho = 0$). In the absence of rotation, the temperature $\beta^{-1}$ increases from $\beta_0$ at infinity to infinite 
values as the black hole outer horizon is approached (i.e.~$r \rightarrow M + \sqrt{M^2 - Q^2}$). 
To better investigate the topology of the horizon structure when $\Omega > 0$, it is convenient 
to cast the equation $\beta^2 = 0$ as:
\begin{equation}
 1 - \frac{1}{\widetilde{r}} + \frac{\widetilde{Q}^2}{\widetilde{r}^2} - \widetilde{\rho}^2 \widetilde{\Omega}^2 = 0,
\end{equation}
where the following notations were introduced:
\begin{equation}
 \widetilde{r} = \frac{r}{2M}, \qquad 
 \widetilde{Q} = \frac{Q}{2M}, \qquad 
 \widetilde{\Omega} = 2M \Omega.
\end{equation}
Figure~\ref{fig:rn_bsq} shows the dependence of $\beta^2 / \beta_0^2$ on $\widetilde{r}$ in the equatorial 
plane $\sin\theta = 1$ at various values of $\widetilde{Q}$ and $\widetilde{\Omega}$. 
In the regions where $\beta^2 > 0$, the local temperature $\beta^{-1}$ is finite and the hydrodynamic 
moments \eqref{eq:hydro_MJ} are well defined. At small enough values of $\widetilde{\Omega}$, the two 
intersections of the graph of
$\beta^2 / \beta_0^2$ with the horizontal axis correspond to the locations of the black hole horizon and 
of the SOL. As $\widetilde{\Omega}$ increases, $\beta^2$ remains negative for all values of $\widetilde{r}$,
showing that the SOL and the black hole event horizon join, forming an exclusion region which incorporates the 
whole equatorial plane. 
It is interesting to note that, at fixed $\widetilde{Q}$, the black hole horizon moves outwards as $\widetilde{\Omega}$
is increased, while the SOL moves inwards, as expected. Figure~\ref{fig:rn_bsq}(d) shows that, for fixed 
$\widetilde{\Omega}$, the black hole horizon moves inwards as $\widetilde{Q}$ is increased, while the SOL is pushed 
outwards, as expected since the charge $Q$ has an inverse effect compared to the mass $M$. 
It is interesting to note that, in the extremal Reissner-Nordstr\"om case, an equilibrium distribution of rotating 
particles sees an event horizon near the black hole which dresses the singularity at $\widetilde{r} = 0$.

The horizon structure of the Reissner-Nordstr\"om spacetime is represented in Fig.~\ref{fig:rn_sol} for various 
values of $\widetilde{Q}$ and $\widetilde{\Omega}$, with the Schwarzschild case ($Q = 0$) shown in Fig.~\ref{fig:rn_sol}(a) 
and the extremal Reissner-Nordstr\"om case shown in Fig.~\ref{fig:rn_sol}(d). It can be seen that
increasing $\widetilde{\Omega}$ at fixed $\widetilde{Q}$ pushes the black hole horizon outwards, while the SOL is 
pulled inwards. At large enough $\widetilde{\Omega}$, these two horizons merge, thus excluding the entire ecuatorial 
plane from the region where $\beta^2 >0$. 

In the absence of rotation, $\zeta$ increases monotonically from $0$ on the event horizon up to $\zeta_0$ as $r\rightarrow \infty$.
In this case, the dependence of $\widetilde{\eta}$ on $r$ is non-monotonic only when $\zeta_0 > \zeta_{\rm max}$, as shown in
Fig.~\ref{fig:rn_eta}(a). The points of maxima $\widetilde{r}_{\rm max}$ which occur outside the outer event horizon are located at:
\begin{equation}
 \widetilde{r}_{\rm max} = 
 \frac{1 + \sqrt{1 - 4\widetilde{Q}^2[1 - (\zeta_{\rm max} / \zeta_0)^2]}}{2[1 - (\zeta_{\rm max} / \zeta_0)^2]},
\end{equation}
valid only for $\zeta_0 > \zeta_{\rm max}$, as shown in Figs.~\ref{fig:rn_eta}(a) and \ref{fig:rn_eta}(b).
When $\zeta_0 < \zeta_{\rm max}$ and $\widetilde{Q} > 0$, the points of maxima are located inside the outer horizon. 

When the rotation is switched on, the location of the points of maxima is given in the general case when $\widetilde{Q} > 0$
by a quartic equation. Figs.~\ref{fig:rn_eta}(c) and \ref{fig:rn_eta}(d) suggest that $\widetilde{\eta}$ can develop 
two points of maxima with a point of local minimum between the event and rotation horizons. For the cases shown 
in these plots, it can be seen that the regime where $\widetilde{\eta}$ presents a local minimum can be obtained 
either by increasing $\widetilde{\Omega}$ at fixed values of $\zeta_0$, or by increasing $\zeta_0$ at fixed values of 
$\widetilde{\Omega}$.

%
%
\section{Conclusion}\label{sec:conc}

In this paper, we employed the tetrad formalism to study the properties of equilibrium states of gases undergoing 
rigid-rotation on spherically-symmetric spacetimes. By employing the Boltzmann equation in conservative 
form \cite{cardall13}, we obtained covariant expressions for the transport coefficients when the Marle 
model for the collision operator is employed. Our results coincide with the expressions on flat spacetime, 
in agreement with the equivalence principle. 
In order to study rigidly-rotating thermal states, we employed a comoving tetrad field, which we obtained by
performing a Lorentz boost on a fixed tetrad which diagonalizes the background spacetime metric. 
Using the tetrad formalism, we obtained expressions for the particle flow 
four-vector and stress-energy tensor corresponding to such states. Furthermore, we discussed the 
formation of speed of light surfaces and their topology in the cases of maximally symmetric spacetimes 
(Minkowski, anti-de Sitter, and de Sitter spaces), as well as in the case of the Reissner-Nordstr\"om black 
hole spacetime (including the Schwarzschild and extremal Reissner-Nordstr\"om cases). 

In constructing the transport coefficients, we considered that the relaxation time is inversely-proportional to 
the average of the M\"oller velocity or the modulus of the velocity. Our analysis showed no qualitative differences
between the results obtained using the two aforementioned definitions for the mean velocity.
We found that the particle number density, energy density,
equilibrium pressure, coefficient of thermal conductivity and coefficient of shear viscosity exhibit a monotonic 
dependence on the inverse temperature $\beta$, such that their properties can be inferred from those of $\beta$.
However, since the coefficient of bulk viscosity $\eta$ attains a maximum value at a finite value of $\beta$, while 
decreasing to $0$ as $\beta$ approaches either $0$ or infinity, its properties were also studied in detail.

For the case of maximally-symmetric spacetimes, we showed that the speed-of-light surface (SOL) forms closer 
to the rotation axis on de Sitter (dS) space compared to Minkowski space, while on anti-de Sitter (adS) space, 
it forms farther away. Furthermore, no SOL forms on adS if 
the rotation parameter $\Omega$ is smaller than the inverse radius of curvature $\omega$. Our analysis 
also revealed that, on AdS, the inverse temperature $\beta$ and all quantities derived from it (i.e.~the stress-energy 
tensor and the transport coefficients) are constant on cones defined by $\Omega \sin \theta = \omega$. 
In particular, $\beta$ is constant throughout the equatorial plane when $\Omega = \omega$.
We found that the coefficient of bulk viscosity can display a non-monotonic behaviour for certain values 
of the relativistic coldness $\zeta_0$ measured at the origin of the spacetime.

In the Reissner-Nordstr\"om case, the SOL plays the role of a ``rotational horizon'', complementing
(and indeed enhancing) the event horizon of the black hole. As the rotation parameter $\Omega$ is 
increased, the distance between the rotational horizon and the rotation axis decreases, while 
the distance between the event horizon and the rotation axis increases. This is also true for the 
case of the extremal Reissner-Nordstr\"om black hole, where the presence of rotation induces an 
event horizon which dresses the singularity at the origin. Increasing the black hole charge 
at fixed rotation parameter has the inverse effect of decreasing the radius of the event horizon,
while pushing the rotational horizon away. When the rotation parameter is non-zero, the coefficient 
of bulk viscosity $\eta$ can exhibit two points of maxima and one local minimum between the 
event horizon and the speed of light surface.

We would like to highlight the fact that the results presented in this paper represent a solid starting 
point for a systematic comparison between kinetic theory results and the properties of rigidly-rotating 
thermal states obtained using quantum field theory on curved spaces. We wish to perform such comparisons 
\cite{ambrus16} for, e.g., rigidly rotating states on the Minkowski spacetime, where analytic results 
are available from the quantum-field theory approach \cite{ambrus14a,becattini15}, as well as from the 
kinetic theory approach \cite{ambrus15}. Furthermore, similar comparisons can be performed for the 
case of the anti-de Sitter space, where analytic results obtained using quantum field theory 
are already available \cite{ambrus14b}. Finally, this work can be extended to the case of 
rigidly-rotating thermal states on axisymmetric space-times, such as the Kerr black hole spacetime.

\begin{acknowledgments}
VEA is indebted to Nistor Nicolaevici and Robert Blaga for useful discussions.
This work was supported by a grant of the Romanian National Authority for Scientific Research and Innovation,
CNCS-UEFISCDI, project number PN-II-RU-TE-2014-4-2910.
\end{acknowledgments}

\appendix

\section{Boltzmann equation with respect to non-holonomic tetrad fields}
\label{app:boltz_tetrad}
In this section of the appendix, the transition from the Boltzmann equation \eqref{eq:boltz} with 
respect to arbitrary coordinates $\{x^\mu\}$ to Eq.~\eqref{eq:boltz_tetrad}, where non-holonomic tetrad 
fields are employed, is presented. Following Ref.~\cite{cercignani02}, it is possible to 
write the exterior derivative of $f$ as follows:
\begin{align}
 df =& \left(\frac{\partial f}{\partial x^\mu}\right)_{p^i} dx^\mu + \frac{\partial f}{\partial p^i} dp^i\nonumber\\
 =& \left(\frac{\partial f}{\partial x^\mu}\right)_{p^\hati} dx^\mu + \frac{\partial f}{\partial p^\hati} dp^\hati,
 \label{eq:df}
\end{align}
where on the first line, $\partial f / \partial x^\mu$ is taken while considering $p^i$ to be constant. On the 
second line, the components $p^\hati = p^\mu \omega_\mu^\hati$ with respect to the tetrad 1-forms $\{\omega^\halpha\}$ 
are kept constant. In order to derive the Boltzmann equation when the components of the momentum 
4-vector are expressed with respect to non-holonomic tetrad fields, the derivatives on the 
first line of Eq.~\eqref{eq:df} must be expressed with respect to derivatives on the second line.

Using Eq.~\eqref{eq:mshell}, 
the following expression can be obtained for the exterior derivative of $p^0$:
\begin{equation}
 dp^0 = -\frac{1}{2} g_{\mu\nu,\lambda} \frac{p^\mu p^\nu}{p_0} dx^\lambda - \frac{p_i}{p_0} dp^i,
\end{equation}
such that the exterior derivative of $p^\hati$ can be written as:
\begin{multline}
 dp^\hati = \left(\frac{\partial \omega^\hati_\nu}{\partial x^\mu} p^\nu - \frac{1}{2} \omega^\hati_0 
 g_{\alpha\beta,\mu} \frac{p^\alpha p^\beta}{p_0}\right) dx^\mu\\
 + \left(\omega^\hati_j - \omega^\hati_0 \frac{p_j}{p_0}\right) dp^j.
\end{multline}
Substituting the above result in Eq.~\eqref{eq:df} yields the following identifications:
\begin{align}
 \left(\frac{\partial f}{\partial x^\mu}\right)_{p^i} =& \left(\frac{\partial f}{\partial x^\mu}\right)_{p^\hati} + 
 \frac{\partial f}{\partial p^\hati} \left(p^\nu \frac{\omega^\hati_\nu}{\partial x^\mu} - 
 \omega^\hati_0 g_{\alpha\beta,\mu} \frac{p^\alpha p^\beta}{2p_0}\right),\nonumber\\
 \frac{\partial f}{\partial p^i} =& \left(\omega^\hatj_i - \omega^\hatj_0 \frac{p_i}{p_0}\right) \frac{\partial f}{\partial p^\hatj}.
\end{align}
The Boltzmann equation can now be written as:
\begin{multline}
 p^\mu \left(\frac{\partial f}{\partial x^\mu}\right)_{p^\hati} - 
 \frac{\partial f}{\partial p^\hati} \left[
 - p^\mu p^\nu \frac{\partial \omega^\hati_\nu}{\partial x^\mu} + 
 \Gamma^j{}_{\mu\nu} p^\mu p^\nu \omega^\hati_j \right.\\
 \left. - \omega^\hati_0 \left(\Gamma^j{}_{\mu\nu} \frac{p^\mu p^\nu p_i}{p_0} - 
 g_{\alpha\beta,\mu} \frac{p^\alpha p^\beta p^\mu}{2p_0}\right)\right] = C[f].
 \label{eq:boltzeq_aux}
\end{multline}
The term involving the derivative of the metric $g_{\alpha\beta,\mu}$ can be written in terms of 
the Christoffel symbols \eqref{eq:christoffel}:
\begin{equation}
 g_{\alpha\beta,\mu} \frac{p^\alpha p^\beta p^\mu}{2p_0} = 
 \Gamma_{\alpha\beta\mu} \frac{p^\alpha p^\beta p^\mu}{p_0},
 \label{eq:boltzeq_aux1}
\end{equation}
while the two terms inside the square bracket on the first line of Eq.~\eqref{eq:boltzeq_aux} can be 
related to the covariant derivative of $\omega^\hatj_\nu$:
\begin{equation}
 - p^\mu p^\nu \frac{\partial \omega^\hati_\nu}{\partial x^\mu} + 
 \Gamma^j{}_{\mu\nu} p^\mu p^\nu \omega^\hati_j = -p^\mu p^\nu \nabla_\mu \omega^\hati_\nu -
 \Gamma^0{}_{\mu\nu} p^\mu p^\nu \omega^\hati_0,
 \label{eq:boltzeq_aux2}
\end{equation}
which can be written in terms of the connection coefficients \eqref{eq:conn_coeff}:
\begin{equation}
 \nabla_\mu \omega^\hatj_\nu = \omega^\hbeta_\mu \nabla_\hbeta \omega^\hatj_\nu 
 = -\Gamma^\hatj{}_{\halpha\hbeta} \omega^\halpha_\nu \omega^\hbeta_\mu.
 \label{eq:boltzeq_aux3}
\end{equation}
Inserting Eqs.~\eqref{eq:boltzeq_aux1}, \eqref{eq:boltzeq_aux2} and \eqref{eq:boltzeq_aux3} 
into Eq.~\eqref{eq:boltzeq_aux} gives the final form for the Boltzmann equation:
\begin{equation}
 p^\halpha e_\halpha^\mu \left(\frac{\partial f}{\partial x^\mu}\right)_{p^\hati} - 
 \Gamma^\hati{}_{\halpha\hbeta} p^\halpha p^\hbeta \frac{\partial f}{\partial p^\hati} = C[f].
\end{equation}

\section{Conservative form of the Boltzmann equation written with respect to non-holonomic tetrad fields}
\label{app:boltz_cons}

In this section of the appendix, we present a derivation of the conservative form \eqref{eq:boltz_cons}
of the Boltzmann equation \eqref{eq:boltz_tetrad}, written with respect to non-holonomic tetrad fields.
Even though the relation between these equations was already found in Ref.~\cite{cardall13}, we present 
this calculation here for completeness.

The term involving the spatial derivatives of $f$ in Eq.~\eqref{eq:boltz_tetrad} can be put in conserivative 
form as follows:
\begin{equation}
 p^\halpha e_\halpha^\mu \frac{\partial f}{\partial x^\mu} = 
 \frac{1}{\sqrt{-g}} \partial_\mu\left(\sqrt{-g} p^\halpha e_\halpha^\mu f\right) -
 \Gamma^\hbeta{}_{\halpha\hbeta} p^\halpha f,\label{eq:boltz_cons_1}
\end{equation}
where the connection coefficient appears from taking the covariant derivative of $e_\halpha^\mu$:
\begin{equation}
 \frac{1}{\sqrt{-g}} \partial_\mu\left(\sqrt{-g} e_\halpha^\mu\right) = 
 \nabla_\mu e_\halpha^\mu = \omega_\mu^\hbeta \Gamma^\hrho{}_{\halpha\hbeta} e_\hrho^\mu
 = \Gamma^\hbeta{}_{\halpha\hbeta}.\label{eq:boltz_cons_1_aux}
\end{equation}

The second term in Eq.~\eqref{eq:boltz_tetrad} can be written as:
\begin{multline}
 \Gamma^\hati{}_{\halpha\hbeta} p^\halpha p^\hbeta \frac{\partial f}{\partial p^\hati} = 
 p^\hatt \frac{\partial}{\partial p^\hati} \left(\Gamma^\hati{}_{\halpha\hbeta} \frac{p^\halpha p^\hbeta}{p^\hatt} f\right) \\
 - fp^\hatt \Gamma^\hati{}_{\halpha\hbeta} \frac{\partial}{\partial p^\hati} \left(\frac{p^\halpha p^\hbeta}{p^\hatt}\right).
 \label{eq:boltz_cons_2_aux}
\end{multline}
The term on the second line in Eq.~\eqref{eq:boltz_cons_2_aux} can be computed as follows.
For the case when the derivative acts on $p^\halpha$, the following expression is obtained:
\begin{subequations}
\begin{align}
 \Gamma^\hati{}_{\halpha\hbeta} \frac{\partial p^\halpha}{\partial p^\hati} =& 
 \Gamma^\hati{}_{\hatt\hbeta} \frac{p_\hati}{p^\hatt} + \Gamma^\hatj{}_{\hatj\hbeta} 
 \label{eq:boltz_cons_2_aux1a}\\
 =& \Gamma^\halpha{}_{\hatt\hbeta} \frac{p_\halpha}{p^\hatt} \label{eq:boltz_cons_2_aux1b}\\
 =& \Gamma^\hatt{}_{\halpha\hbeta} \frac{p^\halpha}{p^\hatt} \label{eq:boltz_cons_2_aux1c},
\end{align}
\end{subequations}
where the term $\Gamma^\hatj{}_{\hatj\hbeta}$ in Eq.~\eqref{eq:boltz_cons_2_aux1a} vanishes 
due to the antisymmetry of the connection coefficients in the first two indices. 
Furthermore, the term $\Gamma^\hati{}_{\hatt\hbeta} p_\hati = \Gamma^\halpha{}_{\hatt\hbeta} p_\halpha$,
since $\Gamma^\hatt{}_{\hatt\hbeta} = 0$. Finally, Eq.~\eqref{eq:boltz_cons_2_aux1c} by noting that 
$\Gamma^\halpha{}_{\hatt\hbeta} p_\halpha = \Gamma_{\halpha\hatt\hbeta} p^\halpha = 
-\Gamma_{\hatt\halpha\hbeta} p^\halpha = \Gamma^\hatt{}_{\halpha\hbeta} p^\halpha$.

Next, the term involving the derivative $p_\hbeta$ can be expressed as:
\begin{subequations}
\begin{align}
 \Gamma^\hati{}_{\halpha\hbeta} \frac{\partial p^\hbeta}{\partial p^\hati} =& 
 \Gamma^\hati{}_{\halpha\hatt} \frac{p_\hati}{p^\hatt} + \Gamma^\hati{}_{\halpha\hati}\label{eq:boltz_cons_2_aux2a}\\
 =& \Gamma_{\hbeta\halpha\hatt} \frac{p^\hbeta}{p^\hatt} + \Gamma^\hbeta{}_{\halpha\hbeta},\label{eq:boltz_cons_2_aux2b}
\end{align}
\end{subequations}
where the relation $\Gamma^\hati{}_{\halpha\hatt} p_\hati = \Gamma_{\hbeta\halpha\hatt} p^\hbeta + 
\Gamma^\hatt{}_{\halpha\hatt} p^\hatt$ was used.

Finally, the term involving the derivative of $p^\hatt$ can be computed as follows:
\begin{subequations}
\begin{align}
 \Gamma^\hati{}_{\halpha\hbeta} \frac{\partial }{\partial p^\hati}\left(\frac{1}{p^\hatt}\right) =&
 -\Gamma^\hati{}_{\halpha\hbeta} \frac{p_\hati}{(p^\hatt)^3} \label{eq:boltz_cons_2_aux3a}\\
 =& -\Gamma_{\hgamma\halpha\hbeta} \frac{p^\hgamma}{(p^\hatt)^3} - \Gamma^\hatt{}_{\halpha\hbeta} \frac{1}{(p^\hatt)^2}.
 \label{eq:boltz_cons_2_aux3b}
\end{align}
\end{subequations}

Inserting Eqs.~\eqref{eq:boltz_cons_2_aux1c}, \eqref{eq:boltz_cons_2_aux2b} and \eqref{eq:boltz_cons_2_aux3b} 
into Eq.~\eqref{eq:boltz_cons_2_aux} yields:
\begin{equation}
 \Gamma^\hati{}_{\halpha\hbeta} p^\halpha p^\hbeta \frac{\partial f}{\partial p^\hati} = 
 p^\hatt \frac{\partial}{\partial p^\hati} \left(\Gamma^\hati{}_{\halpha\hbeta} \frac{p^\halpha p^\hbeta}{p^\hatt} f\right) 
 + \Gamma^\hbeta{}_{\halpha\hbeta} p^\halpha f.
\end{equation}
The final result is obtained by combining the above equation with Eq.~\eqref{eq:boltz_cons_1_aux}:
\begin{equation}
 \frac{1}{\sqrt{-g}} \partial_\mu\left(\sqrt{-g} p^\halpha e_\halpha^\mu f\right) - 
 p^\hatt \frac{\partial}{\partial p^\hati} \left(\Gamma^\hati{}_{\halpha\hbeta} \frac{p^\halpha p^\hbeta}{p^\hatt} f\right) 
 = C[f].
\end{equation}

\end{document}